\documentstyle[preprint,tighten,aps,epsf]{revtex}
\begin{document}

\draft
\title{
Bosonization of interacting fermions in arbitrary dimension beyond the Gaussian
approximation}

\author{Peter Kopietz, Joachim Hermisson and Kurt Sch\H{o}nhammer}
\address{
Institut f\H{u}r Theoretische Physik der Universit\H{a}t G\H{o}ttingen,\\
Bunsenstr.9, D-37073 G\H{o}ttingen, Germany}
\date{February 20, 1995}
\maketitle
\begin{abstract}
We use our recently developed functional bosonization approach
to bosonize interacting fermions in arbitrary dimension $d$
beyond the Gaussian approximation.
Even in $d=1$ the finite curvature of the energy dispersion
at the Fermi surface gives rise to interactions between the bosons.
In higher dimensions scattering processes describing momentum transfer between
different patches on the Fermi surface (around-the-corner processes)
are an additional source for corrections to the Gaussian approximation.
We derive an explicit expression for the leading correction
to the bosonized Hamiltonian and the irreducible self-energy of the bosonic
propagator that takes the finite curvature as well as
around-the-corner processes into account. In the special case that
around-the-corner scattering is negligible, we show that
the self-energy correction to the Gaussian propagator is negligible
if the dimensionless quantities $  ( \frac{q_{c} }{ k_{F}} )^d
F_{0} [ 1 + F_{0} ]^{-1}
\frac{\mu}{\nu^{\alpha}} | \frac{ \partial \nu^{\alpha} }{   \partial \mu} |$
are small compared with unity for all patches $\alpha$.
Here $q_{c}$ is the cutoff of the interaction in wave-vector space, $k_{F}$ is
the
Fermi wave-vector, $\mu$ is the chemical potential,
$F_{0}$ is the  usual dimensionless Landau interaction-parameter,
and $\nu^{\alpha} $ is the {\it{local}} density of states associated
with patch $\alpha$.  We also show that
the well known cancellation between vertex- and self-energy
corrections in one-dimensional systems, which is responsible for the fact that
the
random-phase approximation for the density-density
correlation function is exact in $d=1$, exists also in $d> 1$,
provided
(1) the interaction cutoff $q_{c}$ is small compared with
$k_{F}$, and
(2) the energy dispersion is locally linearized at
the Fermi surface.
Finally, we suggest a new systematic method to calculate corrections to the
RPA, which is based on the perturbative calculation of the
irreducible bosonic self-energy arising from the non-Gaussian terms of
the bosonized Hamiltonian.

\end{abstract}

\narrowtext

\section{Introduction}
\label{sec:intro}

One of the most powerful methods to analyze
systems of interacting fermions in $d=1$ dimension is
the bosonization approach\cite{Tomonaga50}-\cite{Haldane81}. Over the past $30$
years
numerous interesting results have been obtained with this essentially
non-perturbative method,
and the study of one-dimensional fermions via bosonization techniques
continues to be a field of active research.
Because interacting fermions in $d=1$  are not Fermi-liquids\cite{footnote1},
conventional many-body perturbation theory is not applicable to these models.
The success of bosonization in $d=1$ hinges on the
linearization of the energy dispersion at the two Fermi points.
As soon as the curvature of the energy dispersion is taken into account,
bosonization maps  an interacting Fermi system onto an effective interacting
Bose system, which in general cannot be solved exactly.
Of course, one can circumvent
this problem by chosing the energy dispersion of the Fermi system to be linear
{\it{by definition}}\cite{Luttinger63} (Tomonaga-Luttinger model),
but then the physical relevance of this model needs further justification,
because
in realistic materials the energy dispersion is never exactly linear.

In higher dimensions conventional many-body perturbation theory is at least
consistent,
provided the interaction is not too singular\cite{Engelbrecht90}.
Because within the framework of perturbation theory
there is no sign for the breakdown of Fermi-liquid theory,
there was no need to develop non-perturbative methods
for interacting fermions in dimensions higher than one.
Luther's  pioneering attempt to describe higher dimensional Fermi-surface
degrees of
freedom by bosonization was not further pursued\cite{Luther79}.
The discovery of the high-temperature superconductors and Anderson's
suggestion\cite{Anderson90a,Anderson90b} that
the normal state properties of these materials
are a manifestation of non-Fermi-liquid behavior in $d=2$ has revived the
interest to develop non-perturbative methods for
analyzing interacting Fermi systems in $d>1$.
Because of the success of bosonization in $d=1$ it was natural to
further develop this approach in higher dimensions.
Recent progress in this field is based
on ideas due to Haldane\cite{Haldane92,Haldane94},
who realized that Luther's construction\cite{Luther79}  must be generalized
such that
the direction of the net momentum of particle-hole pairs
is not restricted. (In Luther's approach particle-hole pairs carry a net
momentum that
is strictly normal to the local Fermi vector).

\vspace{7mm}

For later reference it is useful to summarize at this point the basic features
of Haldane's "tomographic" construction.
Consider the many-body Hamiltonian
 \begin{eqnarray}
 \hat{H} & = &
 \hat{H}_{0} + \hat{H}_{int}
 \; \; \; ,
 \label{eq:Hdef}
 \\
 \hat{H}_{0} & = & \sum_{\bf{k} } \epsilon_{\bf{k}} \hat{c}^{\dagger}_{\bf{k} }
 \hat{c}_{\bf{k} }
 \; \; \; ,
 \label{eq:H0def}
 \\
 \hat{H}_{int} & = & \frac{1}{2 {{V}}}
 \sum_{  \bf{q} \bf{k}  \bf{k}^{\prime}   }
 f_{\bf{q}}^{\bf{k}  \bf{k}^{\prime}  }
 \hat{c}^{\dagger}_{\bf{k+q} }
 \hat{c}^{\dagger}_{\bf{k^{\prime}-q}  }
 \hat{c}_{\bf{k}^{\prime} }
 \hat{c}_{\bf{k}  }
 \label{eq:H1def}
\; \; \; ,
\end{eqnarray}
where $\hat{c}_{\bf{k} }$ annihilates an electron with wave-vector ${\bf{k}}$,
$V$ is the volume of the system, and
$f_{\bf{q}}^{ \bf{k} \bf{k}^{\prime} }$
are  generalized Landau-parameters that
depend not only on the momentum transfer, but
also on the momenta of the incoming particles.
For simplicity we have suppressed the spin label. The spin
is easily taken into account by defining ${\bf{k}}$ and ${\bf{k}}^{\prime}$ to
be collective labels for wave-vector and spin.
We are implicitly assuming that
only a single band with energy dispersion $\epsilon_{\bf{k}}$ of the
underlying lattice model lies in the vicinity of the Fermi energy, and that
the effect of the other bands is either completely negligible or can be
taken into account via the
definition of effective parameters $\epsilon_{\bf{k}}$ and
$f_{\bf{q}}^{\bf{k}  \bf{k}^{\prime}  }$.
The Fermi surface is the $d-1$-dimensional
hypersurface in ${\bf{k}}$-space  which satisfies
 \begin{equation}
\xi_{\bf{k}} \equiv \epsilon_{\bf{k}} - \mu = 0
\; \; \; ,
 \label{eq:FSdef}
 \end{equation}
where $\mu$ is the chemical potential, measured with respect to the bottom of
the
conduction band.
In $d=1$ the Fermi surface consists of two distinct points $\pm k_{F}$, where
$k_{F}$ is the
Fermi wave-vector, while in $d > 1$ it is a $d-1$-dimensional manifold,
the topology of which depends on the form of $\xi_{\bf{k}}$.
The crucial step in Haldane's construction is the subdivision
of the Fermi surface
into disjoint patches with volume $\Lambda^{d-1}$.
The precise shape of the patches and the size of $\Lambda$ should be chosen
such that within a given patch the curvature of the Fermi surface can
be locally neglected, so that the variations of the local
normal vector can be ignored within a patch.
Note that in this way sufficiently flat Fermi surfaces can be covered with only
a small number of patches.
The extreme case is the Fermi surface of an array of chains in $d=3$
without inter-chain hopping, which consists of two parallel planes
that can be identified with the patches. Obviously, in this case
two patches are sufficient to cover the entire Fermi
surface\cite{Kopietz94b,Schoenhammer94}.

We introduce a label $ \alpha $ to enumerate the patches in some convenient
ordering and
denote the patch labelled $\alpha$ by $\tilde{K}^{\alpha}_{\Lambda}$.
Note that by definition $\tilde{K}^{\alpha}_{\Lambda}$ is
a subset of the Fermi surface, i.e. it is also
a $d-1$-dimensional hypersurface.
To describe all degrees of freedom,
we extend each patch $\tilde{K}^{\alpha}_{\Lambda}$ into a $d$-dimensional
"squat box" $K^{\alpha}_{\Lambda , \lambda  }$ of radial height $\lambda$,
such that the
union $\bigcup_{\alpha} K^{\alpha}_{\Lambda , \lambda }$
agrees with the wave-vector space necessary to label
all degrees of freedom in the system, see Fig.\ref{fig:patch}.
By introducing a radial cutoff $\lambda$, we are
implicitly assuming that all states with wave-vectors outside a shell
with thickness $\lambda$ around the Fermi surface have been integrated out,
and  that the associated finite renormalizations have been taken
into account via the proper definition
of the bare parameters $\epsilon_{\bf{k}}$ and
$f_{\bf{q}}^{ {\bf{k}} {\bf{k}}^{\prime}} $ in Eqs.\ref{eq:H0def} and
\ref{eq:H1def}.
Of course, this construction
is only useful if the physical quantities of interest do not
depend on the choice of the cutoffs $\Lambda$ and $\lambda$.
For sufficiently smooth Fermi surfaces this is indeed the case
as long as we are interested in correlation functions
at separations $ | {\bf{r}} - {\bf{r}}^{\prime} | \gg max \{ \Lambda^{-1} ,
\lambda^{-1} \}$,
and the interaction is dominated by momentum transfers
$|{\bf{q}} | \ll min \{ \Lambda , \lambda \}$\cite{Kopietz95}.
However,
the slightly vague picture of integrating out
the high-energy degrees of freedom without really doing this calculation is
not very useful if we are interested in
the precise numerical value of some physical quantity.
To describe all degrees of freedom, we should therefore
remove the radial cutoff $\lambda$ and work
with boxes $K^{\alpha}_{\Lambda}$ that
are constructed such that
$\bigcup_{\alpha} K^{\alpha}_{\Lambda  }$
covers not only a thin shell around the Fermi surface, but also the
high-energy degrees of freedom.
Of course, if we would like to linearize the energy dispersion around the Fermi
surface, we
should keep the radial cutoff finite, because the linearization is
only good close to the Fermi surface.
However, our bosonization approach does not rely on the linearization, so that
in some cases we can remove the radial cutoff and can dispose of the picture
of having integrated out the high-energy degrees of freedom.
For example, to discuss the  homogeneous electron gas in $d=2$
we take $\epsilon_{\bf{k}} = \frac{ {\bf{k}}^2}{2 m }$,
$f_{\bf{q}}^{ {\bf{k}} {\bf{k}}^{\prime} }
= \frac{ 2 \pi e^2}{ |{\bf{q}}| }$, and choose
the boxes $K^{\alpha}_{\Lambda}$
shown in Fig.\ref{fig:patch2}.
We shall come back to the cutoff problem in the radial direction in
Sec.\ref{subsec:hidden}.

Because the union of all $K^{\alpha}$
agrees by construction with
the total relevant ${\bf{k}}$-space, we have
 \begin{equation}
 \sum_{\alpha} \Theta^{\alpha} ( {\bf{k}} ) = 1
 \label{eq:Thetasum}
 \; \; \; ,
 \end{equation}
where
 \begin{equation}
 \Theta^{\alpha} ( {\bf{k}} ) =
 \left\{
 \begin{array}{cl}
 1 & \mbox{ if ${\bf{k}} \in K^{\alpha}$ } \\
 0 & \mbox{ else }
 \end{array}
 \right.
 \; \; \; .
 \label{eq:Thetadef}
 \end{equation}
For simplicity we have omitted the cutoff label of $K^{\alpha}$.
In the relevant Hilbert space
Eq.\ref{eq:Thetasum}
is correct for both type of boxes $K^{\alpha}_{\Lambda}$ and
$K^{\alpha}_{\Lambda , \lambda}$ discussed above.
Let us denote by ${\bf{k}}^{\alpha}$ a vector on the Fermi surface
($\xi_{ {\bf{k}}^{\alpha} }= 0 $)
that points
to the (suitably defined) center of patch $\alpha$.
The set of vectors $\left\{ {\bf{k}}^{\alpha} \right\}$ define
the origins of local coordinate systems on the Fermi surface. Such a collection
of coordinate systems is also called an {\it{atlas}}.
Let us now measure the wave-vector ${\bf{k}}$
locally with respect to the origin of the appropriate coordinate system.
Using Eq.\ref{eq:Thetasum}, we may write
 \begin{equation}
 \xi_{\bf{k}}
 = \sum_{\alpha}
 \Theta^{\alpha} ( {\bf{k}} )
  \xi_{\bf{k}}
 =
 \sum_{\alpha}
 \Theta^{\alpha} ( {\bf{k}} )
 \xi^{\alpha}_{ {\bf{k}} - {\bf{k}}^{\alpha}  }
 \label{eq:partition}
 \; \; \; ,
 \end{equation}
where the quantity $\xi^{\alpha}_{\bf{q}}$ is defined by
 \begin{equation}
 \xi^{\alpha}_{\bf{q}}  \equiv \xi_{ {\bf{k}}^{\alpha} + {\bf{q}}}
 \; \; \; .
  \label{eq:xialphaq}
 \end{equation}
The crucial step in the conventional bosonization approach is
the local linearization of the energy dispersion within a given patch,
 \begin{equation}
 \xi^{\alpha}_{\bf{q}} =  {\bf{v}}^{\alpha} \cdot  {\bf{q}}  + O ( {\bf{q}}^2 )
 \label{eq:xilin}
 \; \; \; ,
 \end{equation}
where
 \begin{equation}
 {\bf{v}}^{\alpha} =  \nabla_{\bf{k}} \left. \epsilon_{\bf{k}}
 \right|_{ {\bf{k}} = {\bf{k}}^{\alpha} }
 \label{eq:valphadef}
 \end{equation}
is the local Fermi velocity at patch $\alpha$,
and we measure frequencies in units of energy, which amounts to
formally setting $\hbar = 1$.
Note that such a linearization would not be possible
in a rigid coordinate system.
It is important to stress that this linearization should be made around the
true chemical potential of the interacting many-body system\cite{Anderson93},
and that
${\bf{v}}^{\alpha}$ should be identified with the renormalized local
Fermi velocity, taking
effective mass renormalizations into account\cite{Kopietz95}.

The operator-approach to bosonization follows then closely the
usual construction for one-dimensional systems\cite{Haldane81}.
This approach has been further developed by
Houghton, Kwon, Marston and Shankar\cite{Houghton93}-\cite{Kwon95}, and
independently
by Castro Neto and Fradkin\cite{Castro94a}-\cite{Castro94c}.
The fundamental objects are
the local density operators associated with the patches,
 \begin{equation}
 \hat{\rho}^{\alpha}_{\bf{q}} = \sum_{\bf{k}} \Theta^{\alpha} ( {\bf{k}} )
 \hat{c}^{\dagger}_{\bf{k}} \hat{c}_{\bf{k} + \bf{q} }
 \label{eq:rhoopdef}
 \; \; \; .
 \end{equation}
In terms of these operators  the
interaction-part of $\hat{H}$ can be written as
 \begin{equation}
 \hat{H}_{int} = \frac{ 1}{2V} \sum_{\bf{q}}  \sum_{ \alpha \alpha^{\prime} }
 {{f}}_{\bf{q}}^{\alpha \alpha^{\prime}}
 : \hat{\rho}_{ - \bf{q}}^{\alpha} \hat{\rho}_{\bf{q}}^{\alpha^{\prime}} :
 \label{eq:Hintcourse}
 \; \; \; ,
 \end{equation}
where $: \ldots :$ denotes normal ordering,
and it is assumed that the variations of $f^{ {\bf{k}} {\bf{k}}^{\prime}
}_{\bf{q}}$
are negligible if
${\bf{k}}$ and ${\bf{k}}^{\prime}$ are restricted to given patches,
so that it is allowed to introduce the coarse-grained interaction function
 \begin{equation}
 {{f}}_{\bf{q}}^{ \alpha \alpha^{\prime}} =
 \frac{ \sum_{ {\bf{k}} {\bf{k}}^{\prime}  }
  \Theta^{\alpha} ( {\bf{k}} )
  \Theta^{\alpha^{\prime} } ( {\bf{k}}^{\prime} )
  f^{ \bf{k k^{\prime}} }_{\bf{q}}
  }{
  \sum_{ {\bf{k}}  {\bf{k}}^{\prime} }
  \Theta^{\alpha} ( {\bf{k}} )
  \Theta^{\alpha^{\prime} } ( {\bf{k}}^{\prime} )
  }
   \label{eq:Landaufuncdef}
   \; \; \; .
 \end{equation}
By considering the commutation relations of the operators
$\hat{\rho}^{\alpha}_{\bf{q}}$ with each other and with $\hat{H}$ in the
restricted Hilbert space
of states with wave-vectors close to the Fermi surface, it is possible to show
that
the $\hat{\rho}^{\alpha}_{\bf{q}}$ {\it{approximately}} obey bosonic
commutation relations, and that
$\hat{H}$ is {\it{approximately}} quadratic in the local density operators, so
that it
is exactly solvable\cite{Haldane94},\cite{Houghton93}-\cite{Castro94c}.

It should be mentioned that an alternative non-perturbative approach to
study interacting fermions in dimensions higher than one has been
advanced by Castellani, Di Castro and Metzner\cite{Castellani94}.
This method is based on Ward identities and leads to results that are
very similar to those of Refs.\cite{Houghton93}-\cite{Castro94c}.

\vspace{7mm}

Recently two of us have developed a functional integral method to bosonize
interacting Fermi systems in arbitrary dimensions\cite{Kopietz94},
which avoids the algebraic manipulations of commutators that is necessary in
the operator approach.
A similar method has been developed independently by Fr\H{o}hlich and
collaborators\cite{Frohlich94}, although in their work the
physical content is somehow hidden in elegant mathematics.
In the context of the one-dimensional Tomonaga-Luttinger model
the functional bosonization technique has first been discussed by
Fogedby\cite{Fogedby76}, and later by Lee and Chen\cite{Lee88}.
Besides reproducing all of the results of
Refs.\cite{Houghton93}-\cite{Castro94c}
with a minimum of algebra\cite{Kopietz95,Kopietz94}, our functional
bosonization  approach
has several other advantages. First of all, it can cope with
retardation and non-locality in a very simple way, because path integrals are
the natural
language to describe these important many-body effects.
Hence, the inclusion of electron-phonon interactions or the treatment of the
retarded
interaction mediated by transverse gauge-fields is straightforward in our
approach\cite{Kopietz95}.
A second important advantage of our approach is that it opens the way
for a controlled calculation of corrections to the non-interacting
boson approximation. As already mentioned, these non-Gaussian corrections exist
in any dimension if the energy dispersion is not linearized at the Fermi
surface.
Some time ago Haldane pointed out
in his seminal article on one-dimensional interacting Fermi
systems\cite{Haldane81}
that the perturbation theory for the effective interacting boson-problem,
which arises from the bosonization mapping, is most likely well behaved.
In this paper we shall verify this hypothesis not only in $d=1$,
but calculate explicitly
the leading correction to the non-interacting boson approximation
in arbitrary dimension.
We shall show that this correction is closely related
to the local field correction to the random-phase approximation (RPA) for
the density-density correlation function.

The rest of this paper is organized as follows:
In Sec.\ref{sec:densbos} we shall give a detailed description
of our functional bosonization technique, and develop the general formalism
that is necessary to calculate the corrections to the non-interacting boson
approximation.
Sec.\ref{sec:Gauss} contains the calculation of the
bosonized Hamiltonian and the density-density correlation function
within the Gaussian approximation.
The calculation of the leading correction to the
Gaussian approximation will be presented in Sec.\ref{sec:beyond}.
We show that the non-Gaussian correction terms to the
bosonized Hamiltonian correspond to the local field corrections to the RPA,
and calculate the hidden small
parameter which determines the range of validity of the Gaussian approximation.
Finally, in Sec.\ref{sec:conclusions} we present our conclusions and
discuss some open problems.

\section{General formalism of functional bosonization}
\label{sec:densbos}
\setcounter{equation}{0}

In this section we shall describe our functional bosonization approach and
prove that
under certain conditions there exists for all $d$ a remarkable cancellation
between self-energy and vertex corrections, which implies that the Gaussian
approximation
becomes exact.

\subsection{
The effective action for collective density fluctuations}

It is convenient to derive the bosonized Hamiltonian by
first considering the imaginary-frequency  density-density
correlation function of our interacting many-body system,
which is defined by
 \begin{equation}
 \Pi ( {\bf{q}} , i \omega_{m} ) \equiv \Pi ( q ) =
 \frac{1}{\beta V } \int_{0}^{\beta}  d \tau
 \int_{0}^{\beta} d \tau^{\prime}
 e^{ - i \omega_{m}  ( \tau - \tau^{\prime}) }
 < {\cal{T}} \left[ \hat{\rho}_{\bf{q}} ( \tau )
 \hat{\rho}_{ -{\bf{q} } }  ( \tau^{\prime} ) \right] >
 \label{eq:cordens}
 \; \; \; ,
 \end{equation}
where $\beta = 1/T$ is the inverse temperature,
$\hat{\rho}_{\bf{q}} = \sum_{\bf{k}} \hat{c}^{\dagger}_{\bf{k}}
\hat{c}_{\bf{k+q}}$ is the
operator representing the Fourier transform of the total density, and
${\cal{T}}$ denotes time-ordering
in imaginary time.
Throughout this work $q = [ {\bf{q}} , i \omega_{m} ]$ is a collective label
for
wave-vector ${\bf{q}}$ and bosonic Matsubara frequency $\omega_{m} = 2 \pi m
T$.
The average in Eq.\ref{eq:cordens} denotes thermal average with respect to all
degrees of freedom
in the system.
An essential first step is the decomposition of
Eq.\ref{eq:cordens} into
contributions from the various patches.
Using Eq.\ref{eq:Thetasum}, we may write
 \begin{equation}
 \Pi ( q) = \sum_{\alpha \alpha^{\prime}} \Pi^{\alpha \alpha^{\prime}} ( q )
 \label{eq:Ptotdecompose}
 \; \; \; ,
 \end{equation}
where
 \begin{equation}
 \Pi^{\alpha  \alpha^{\prime} } ( q )
 = \frac{1}{\beta V } \int_{0}^{\beta}  d \tau
 \int_{0}^{\beta} d \tau^{\prime}
 e^{ - i \omega_{m}  ( \tau - \tau^{\prime}) }
 < {\cal{T}} \left[ \hat{\rho}^{\alpha}_{\bf{q}} ( \tau )
 \hat{\rho}^{ \alpha^{\prime} }_{ -{\bf{q} } } ( \tau^{\prime} ) \right] >
 \label{eq:cordensloc}
 \; \; \; .
 \end{equation}
We shall refer to $\Pi ( q )$ as the {\it{global}} density-density correlation
function,
and to $\Pi^{\alpha \alpha^{\prime}} ( q )$ as the {\it{patch}} density-density
correlation
function.
$\Pi^{\alpha \alpha^{\prime}} ( q )$
can be represented as a functional integral
over a Grassmann-field $\psi$ and an auxiliary field $ \phi_{ q }^{\alpha}$
that mediates the interaction between the patch densities
\cite{Kopietz95,Kopietz94,Popov87,Negele88},
 \begin{equation}
 \Pi^{\alpha  \alpha^{\prime} } ( q )
 = \frac{\beta}{V}
 \frac{
 \int {\cal{D}} \left\{ \psi \right\}
 {\cal{D}} \left\{ \phi^{\alpha} \right\}
 \rho^{\alpha}_{-q} \rho^{\alpha^{\prime}}_{q}
 e^{- S_{1} \left\{ \psi \right\} - {S}_{2} \left\{ \psi , \phi^{\alpha}
\right\}
 - {S}_{3} \left\{ \phi^{\alpha} \right\} }
}
 { \int {\cal{D}} \left\{ \psi \right\}
 {\cal{D}} \left\{ \phi^{\alpha} \right\}
 e^{- S_{1} \left\{ \psi \right\} - {S}_{2} \left\{ \psi , \phi^{\alpha}
\right\}
 - {S}_{3} \left\{ \phi^{\alpha} \right\} } }
 \label{eq:cordensfunc}
 \; \; \; ,
 \end{equation}
where $\rho^{\alpha}_{q}$ is now a quadratic functional of the
Grassmann-fields,
 \begin{equation}
 \rho_{q}^{\alpha}  =  \sum_{k}
 \Theta^{\alpha} ( {\bf{k}} )
 \psi^{\dagger}_{k} \psi_{k+q}
 \label{eq:rhoalpha}
 \; \; \; .
 \end{equation}
Here $\psi_{k} $ is the Fourier component of the Grassmann-field,
where $k$ is again a collective label for wave-vector ${\bf{k}}$ and fermionic
Matsubara frequency $\tilde{\omega}_n = 2 \pi ( n + \frac{1}{2} ) T$.
The three contributions to the action in Eq.\ref{eq:cordensfunc} are
 \begin{eqnarray}
 S_{1} \left\{ \psi \right\}
 & = & \beta \sum_{ k } \psi^{\dagger}_{ k } \left[ -i \tilde{\omega}_{n} +
 \xi_{\bf{k}}  \right] \psi_{ k }
 \; \; \; ,
 \label{eq:S1Fourier}
 \\
 {S}_{2} \left\{ \psi , {{\phi}}^{\alpha} \right\}
 & = &  i \sum_{ q } \sum_{\alpha}
  \phi^{\alpha}_{-q} \rho^{\alpha}_{q}
  \; \; \; ,
 \label{eq:S2Fourieralphaphi}
 \\
 {S}_{3} \left\{ {{\phi}}^{\alpha}  \right\} & = &
 \frac{1}{2} \sum_{q} \sum_{\alpha \alpha^{\prime} }
 [ \underline{\tilde{f}}_{ {{q}} }^{-1} ]^{ \alpha \alpha^{\prime} }
 \phi_{-q}^{\alpha} \phi_{q}^{\alpha^{\prime}}
 \label{eq:S3Fourieralphaphi}
 \; \; \; ,
 \end{eqnarray}
where $\underline{\tilde{f}}_{q}$ is a dimensionless matrix in the
patch labels, which is defined
in terms of the coarse-grained Landau parameters $f_{\bf{q}}^{\alpha
\alpha^{\prime}}$ of
Eq.\ref{eq:Landaufuncdef} via
 \begin{equation}
 \frac{V}{\beta} [ \underline{\tilde{f}}_{q} ]^{\alpha \alpha^{\prime} } =
f_{\bf{q}}^{\alpha \alpha^{\prime}}
 \label{eq:ftildedef}
 \; \; \; .
 \end{equation}
The field $\phi^{\alpha}_{q}$ can be considered as the dual field of the
composite fermionic field $\rho^{\alpha}_{q}$\cite{Kopietz94,Frohlich94}.
We have intentionally chosen our notation for the actions $S_{1}$, $S_{2}$ and
$S_{3}$ such that it matches the notation used by Feynman and
Hibbs\cite{Feynman65}:
$S_{1} $ is the action for the matter degrees of freedom,
$S_{2}$ describes the coupling between matter and a bosonic field
$\phi^{\alpha}_{q}$ that mediates
the interaction between the matter, and
$S_{3}$ is the action of the bosonic field.
In the case of the three dimensional Coulomb interaction
the field $\phi^{\alpha}_{q}$ can be identified with the scalar potential of
electromagnetism in the Coulomb gauge, and it is straightforward to include
also the transverse radiation field in our formalism\cite{Kopietz95}.

In order to introduce a collective bosonic field that represents the bosonized
local density, we decouple
 ${S}_{3} \left\{ {{\phi}}^{\alpha}  \right\}$ by means of another
Hubbard-Stratonowich transformation\cite{Kopietz95,Kopietz94,Lee88}.
The integration over the Grassmann-field $\psi$ is then quadratic and can be
formally
carried out, so that we can
eliminate the composite fermionic field $\rho^{\alpha}_{q}$ in favour of a
collective bosonic field $\tilde{\rho}^{\alpha}_{q}$.
We arrive at the following  exact expression for
the patch density-density correlation function,
 \begin{equation}
 \Pi^{\alpha  \alpha^{\prime} } ( q )
 = \frac{\beta}{V}
 \frac{
 \int
 {\cal{D}} \left\{ \tilde{\rho}^{\alpha} \right\}
 e^{- {S}_{int} \left\{ \tilde{\rho}^{\alpha} \right\} }
 \tilde{\rho}^{\alpha}_{-q} \tilde{\rho}^{\alpha^{\prime}}_{q}
 \int {\cal{D}} \left\{ \phi^{\alpha} \right\}
 e^{
  i   \sum_{q \alpha} \phi^{\alpha}_{-q} \tilde{\rho}^{\alpha}_{q} }
 e^{ - S_{kin} \left\{ \phi^{\alpha} \right\}  }
}
 { \int
 {\cal{D}} \left\{ \tilde{\rho}^{\alpha} \right\}
 e^{- {S}_{int} \left\{ \tilde{\rho}^{\alpha} \right\} }
 \int {\cal{D}} \left\{ \phi^{\alpha} \right\}
 e^{
  i   \sum_{q \alpha} \phi^{\alpha}_{-q} \tilde{\rho}^{\alpha}_{q} }
 e^{ - S_{kin} \left\{ \phi^{\alpha} \right\}  }
 }
 \label{eq:Piphirho}
 \; \; \; ,
 \end{equation}
where
 \begin{eqnarray}
 {S}_{int} \left\{ \tilde{\rho}^{\alpha} \right\}  & = &
 \frac{1}{2} \sum_{q} \sum_{\alpha \alpha^{\prime}}
 [ \underline{\tilde{f}}_{q} ]^{ \alpha \alpha^{\prime} }
 \tilde{\rho}^{\alpha}_{-q} \tilde{\rho}^{\alpha^{\prime}}_{q}
 \; \; \; ,
 \label{eq:Sintrhodef}
 \\
 {S}_{kin} \left\{ \phi^{\alpha} \right\}
  & = & - Tr \ln \left[ \hat{1} - \hat{G}_{0} \hat{V} \right]
 \label{eq:Skintracelog}
 \; \; \; .
 \end{eqnarray}
Here the trace is over all wave-vectors
and frequencies, and $\hat{G}_{0}$ and $\hat{V}$ are infinite matrices in
wave-vector and
frequency space, with matrix elements given by
 \begin{equation}
 [ \hat{G}_{0} ]_{ k k^{\prime} }
   =  \delta_{k k^{\prime} } G_{0} ( k)
   \; \; \; , \; \; \;
 G_{0} ( k ) = \frac{1}{ i \tilde{\omega}_{n} - \xi_{\bf{k}} }
  \; \; \; ,
 \label{eq:hatG0}
 \end{equation}
 \begin{equation}
 [ \hat{V} ]_{ k k^{\prime} }
  =    \frac{i}{\beta} \sum_{\alpha} \Theta^{\alpha} ( {\bf{k}} )
  {\phi}^{\alpha}_{k- k^{\prime}}
 \label{eq:hatVphi}
 \; \; \; .
 \end{equation}
The subscript on $S_{ kin } \left\{ \phi^{\alpha} \right\} $ indicates that
this quantity is closely related to the bosonized kinetic energy.
To be precise, below we shall show that the bosonized kinetic energy is
given by the negative logarithm of the functional-
Fourier transform of $e^{ - S_{kin} \left\{ \phi^{\alpha} \right\}}$,
i.e.
 \begin{equation}
 S_{kin} \left\{ \tilde{\rho}^{\alpha} \right\}
 = - \ln \left[
 \int {\cal{D}} \left\{ \phi^{\alpha} \right\}
 e^{
  i   \sum_{q \alpha} \phi^{\alpha}_{-q} \tilde{\rho}^{\alpha}_{q} }
 e^{ - S_{kin} \left\{ \phi^{\alpha} \right\}  }
 \right]
 \label{eq:Skinrhodef}
 \; \; \; .
 \end{equation}
With this definition Eq.\ref{eq:Piphirho} can also be written as
 \begin{equation}
 \Pi^{\alpha  \alpha^{\prime} } ( q )
 = \frac{\beta}{V}
 \frac{
 \int
 {\cal{D}} \left\{ \tilde{\rho}^{\alpha} \right\}
 e^{- {S}_{eff} \left\{ \tilde{\rho}^{\alpha} \right\} }
 \tilde{\rho}^{\alpha}_{-q} \tilde{\rho}^{\alpha^{\prime}}_{q}
}
 { \int
 {\cal{D}} \left\{ \tilde{\rho}^{\alpha} \right\}
 e^{- {S}_{eff} \left\{ \tilde{\rho}^{\alpha} \right\} }
 }
 \label{eq:Pirho}
 \; \; \; ,
 \end{equation}
with
 \begin{equation}
 S_{eff} \left\{ \tilde{\rho}^{\alpha} \right\}
 =
 S_{int} \left\{ \tilde{\rho}^{\alpha} \right\}
 + S_{kin} \left\{ \tilde{\rho}^{\alpha} \right\}
 \label{eq:Seffrhodef}
 \; \; \; .
 \end{equation}
Note that
the electron-electron interaction has been treated exactly,
while the bosonized kinetic $S_{kin} \left\{ \tilde{\rho}^{\alpha} \right\}$
can in general only be calculated approximately.
The calculation consists of two steps. First of all, one should
calculate the functional $S_{kin} \left\{ \phi^{\alpha} \right\}$.
In praxis, this has to be done by expanding the logarithm
in Eq.\ref{eq:Skintracelog} and performing the traces
in each term separately,
 \begin{eqnarray}
 S_{kin}
 \left\{ \phi^{\alpha} \right\}
 & = & \sum_{n=1}^{\infty}
 S_{kin , n }
 \left\{ \phi^{\alpha} \right\}
 \; \; \; ,
 \label{eq:Skinexp}
 \\
 S_{kin , n }
 \left\{ \phi^{\alpha} \right\}
 & = & \frac{1}{n} Tr \left[ \hat{G}_{0} \hat{V} \right]^n
 \; \; \; .
 \label{eq:tracelogexp}
 \end{eqnarray}
To obtain the bosonized kinetic energy $S_{kin} \left\{ \tilde{\rho}^{\alpha}
\right\}$
it is necessary to
perform the functional Fourier transformation in Eq.\ref{eq:Skinrhodef}.
Of course, in general also this step has to be carried out perturbatively.
If we ignore all terms with $n \geq 3$ in the expansion \ref{eq:Skinexp},
we obtain the Gaussian approximation.
In Ref.\cite{Kopietz94} we have pointed out that in the
one-dimensional
Tomonaga-Luttinger model all higher order terms
vanish identically, so that we have exactly
 \begin{equation}
   - Tr \ln \left[ \hat{1} - \hat{G}_{0} \hat{V} \right]
  = Tr \left[ \hat{G}_{0} \hat{V} \right]
  + \frac{1}{2}
  Tr \left[ \hat{G}_{0} \hat{V} \right]^2
  \label{eq:Trlogexact}
  \; \; \; .
  \end{equation}
The functional Fourier transformation of
$S_{kin} \{ \phi^{\alpha} \}$
involves
then a simple Gaussian integral, so that the bosonized kinetic
energy can be calculated exactly.
In the next subsection we shall show that
in a certain physically relevant limit
the remarkable cancellation between
self-energy and vertex corrections that is responsible for the
validity of Eq.\ref{eq:Trlogexact} in $d=1$ happens also in higher dimensions.

\subsection{Generalized closed loop theorem}
\label{subsec:closedloop}

Graphically, the trace in Eq.\ref{eq:tracelogexp} can be represented as a
closed
fermion loop with $n$ external $\phi^{\alpha}$-fields, see
Fig.\ref{fig:closedloop}.
Performing the
trace in  Eq.\ref{eq:tracelogexp} we obtain
 \begin{equation}
 S_{kin,n} \left\{ { { {\phi}^{\alpha}}} \right\}  =
 \frac{1}{n} \sum_{q_{1}  \ldots q_{n} }
 \sum_{\alpha_{1} \ldots \alpha_{n} }
 U_{n} (
 q_1 \alpha_{1} \ldots q_{n} \alpha_{n}  ) {\phi}^{\alpha_{1}}_{q_{1}} \cdots
 {\phi}^{\alpha_{n}}_{q_{n}}
 \; \; \; ,
 \label{eq:Seffphin}
 \end{equation}
where the dimensionless vertices $U_{n}$ are given by
 \begin{eqnarray}
 U_{n} ( q_1  \alpha_{1}  \ldots q_{n} \alpha_{n}  )
 & = &
\delta_{ {q}_{1} + \ldots
+ {{q}}_{n} , 0 }
 \left( \frac{i}{\beta} \right)^n
 \frac{1}{n!} \sum_{P(1 \ldots n)} \sum_{k}
 \Theta^{\alpha_{P_1}} ( {\bf{k}} )
 \nonumber
 \\
 & \times &
 \Theta^{\alpha_{P_2}} ( {\bf{k}} + {\bf{q}}_{P_1} )
 \cdots
 \Theta^{\alpha_{P_n}} ( {\bf{k}} + {\bf{q}}_{P_1} + \ldots +
{\bf{q}}_{P_{n-1}} )
 \nonumber
 \\
 & \times  &
 G_{0} ( k )
 G_{0} ( k + q_{P_1} ) \cdots
 G_{0} ( k + q_{P_1} + \ldots + q_{P_{n-1}} )
 \; .
 \label{eq:Uvertex}
 \end{eqnarray}
Here $\delta_{ {q}_{1} + \ldots + q_{n} , 0 }$ denotes a
Kronecker-$\delta$ in wave-vector and frequency space.
We have used the invariance of $S_{kin, n} \left\{ {\phi}^{\alpha} \right\}$
under relabeling of the fields to symmetrize the vertices $U_{ n }$
with respect to the interchange of any two
labels. The sum $\sum_{P(1 \ldots n)}$ is over the $n!$ permutations of $n$
integers, and
$P_{i}$ denotes the image of $i$ under the permutation.
Note that the vertices $U_{n}$ are
uniquely determined by the energy dispersion
$\epsilon_{\bf{k}} - \mu $
that defines the free fermionic action.
The amazing point is now that there exists a physically interesting limit where
all vertices $U_{n}$ with $n \geq 3$ vanish.
This limit is characterized by the requirement that the following two
approximations $(A1)$ and $(A2)$
become accurate:

\vspace{7mm}

{\it{
(A1): High density-limit or small momentum-transfer limit:}}
Suppose that there exists
a cutoff $q_{c} \ll  k_{F}$ such
that the contribution from fields $\phi^{\alpha}_{q}$
with $|{\bf{q}} | \geq q_{c}$ to physical observables becomes
negligibly small.
Because the fields $\phi^{\alpha}_{q}$ mediate the interaction
between the fermions, this
condition is equivalent with  the requirement that the
nature of the {\it{bare interaction}}
$\underline{\tilde{f}}_{q}$ should be such that the resulting {\it{effective
screened interaction}}
(which takes into account the modification of the bare interaction
between two particles due to the presence of all other particles)
is negligibly small for $|{\bf{q}} | \geq q_{c}$.
If this condition is satisfied, we
may approximate in Eq.\ref{eq:Uvertex}
 \begin{eqnarray}
 \lefteqn{\Theta^{\alpha_{P_1}} ( {\bf{k}} )
 \Theta^{\alpha_{P_2}} ( {\bf{k}} + {\bf{q}}_{P_1} )
 \cdots
 \Theta^{\alpha_{P_n}} ( {\bf{k}} + {\bf{q}}_{P_1} + \ldots +
{\bf{q}}_{P_{n-1}} ) }
 \nonumber
 \\
 & &
 \approx
 \delta^{ \alpha_{P_{1}} \alpha_{P_{2}} }
 \delta^{ \alpha_{P_{1}} \alpha_{P_{3}} }
 \cdots
 \delta^{ \alpha_{P_{1}} \alpha_{P_{n}} }
 \Theta^{\alpha_{P_1}} ( {\bf{k}} )
 \label{eq:highdens}
 \; \; \; ,
 \end{eqnarray}
because the ${\bf{k}}$-sum in Eq.\ref{eq:Uvertex} is
dominated by wave-vectors of the order of $k_{F}$.
This approximation is correct to leading order in
$q_{c} / k_{F}$, and becomes exact in the limit
$q_{c} / k_{F} \rightarrow 0$.
Note that this limit is approached either at high densities,
where $k_{F} \rightarrow \infty$ at constant $q_{c}$, or in the limit
that the range $q_{c}$ of the effective interaction in momentum space
approaches zero while $k_{F}$ is held constant.
It follows that up to higher order corrections in $q_{c} / k_{F}$  the vertex
 $U_{n} ( q_1  \alpha_{1}  \ldots q_{n} \alpha_{n}  ) $ is diagonal in all
 patch labels,
 \begin{equation}
 U_{n} ( q_1  \alpha_{1}  \ldots q_{n} \alpha_{n}  )
 = \delta^{\alpha_{1} \alpha_{2} } \cdots \delta^{\alpha_{1} \alpha_{n}}
 U_{n}^{\alpha_{1}} ( q_{1} \ldots q_{n} )
 \label{eq:Undiag}
 \; \; \; ,
 \end{equation}
with
 \begin{eqnarray}
 U_{n}^{\alpha} ( q_{1} \ldots q_{n} )
 &  = &
\delta_{ {q}_{1} + \ldots
+ {{q}}_{n} , 0 }
 \left( \frac{i}{\beta} \right)^n
 \frac{1}{n!} \sum_{P(1 \ldots n)} \sum_{k}
 \Theta^{\alpha} ( {\bf{k}} )
 \nonumber
 \\
 & \times &
 G_{0} (k)
 G_{0} ( k + q_{P_1} ) \cdots
 G_{0} ( k + q_{P_1} + \ldots + q_{P_{n-1}} )
 \; \; \; .
 \label{eq:Uvertexdiag}
 \end{eqnarray}
Below we shall refer to the approximation \ref{eq:highdens} as the
{\it{diagonal-patch approximation}}.
It is important to note that at finite $q_{c} / k_{F}$ this approximation
can only become exact  in $d=1$, because in this case the Fermi surface
consists of two widely separated points.
Except for special cases\cite{Kopietz94b,Schoenhammer94,Kopietz95},
the covering of the Fermi surface in $d> 1$ involves at least some adjacent
patches,
so that in higher dimensions there exist always patches which can be connected
by an arbitrarily small
momentum transfer ${\bf{q}}$.
These "around-the-corner" processes are ignored within the diagonal-patch
approximation $(A1)$.

\vspace{7mm}

{\it{
(A2): Local linearization of the energy dispersion at the Fermi-surface:}}
Suppose we linearize the energy dispersion locally
by approximating $\xi^{\alpha}_{\bf{q}} \approx  {\bf{v}}^{\alpha} \cdot
{\bf{q}}$,
see Eqs.\ref{eq:xilin} and \ref{eq:valphadef}.
Shifting the summation wave-vector in Eq.\ref{eq:Uvertexdiag} according to
${\bf{k}} = {\bf{k}}^{\alpha} + {\bf{q}}$ and defining the
linearized Greens-function
 \begin{equation}
 G^{\alpha}_{0} (q) \equiv G_{0}^{\alpha} ( {\bf{q}} , i \tilde{\omega}_{n} )
 = \frac{1}{ i \tilde{\omega}_{n} -  {\bf{v}}^{\alpha} \cdot {\bf{q}} }
 \label{eq:Galphalin}
 \; \; \; ,
 \end{equation}
Eq.\ref{eq:Uvertexdiag} reduces to
 \begin{eqnarray}
 U_{n}^{\alpha} ( q_{1} \ldots q_{n} )
  & =  &
\delta_{ {q}_{1} + \ldots
+ {{q}}_{n} , 0 }
 \left( \frac{i}{\beta} \right)^n
 \frac{1}{n!} \sum_{P(1 \ldots n)} \sum_{q}
 \Theta^{\alpha} ( {\bf{k}}^{\alpha} + {\bf{q}} )
 \nonumber
 \\
 & \times &
 G_{0}^{\alpha} (q)
 G_{0}^{\alpha} ( q + q_{P_1} ) \cdots
 G_{0}^{\alpha} ( q + q_{P_1} + \ldots + q_{P_{n-1}} )
 \; \; \; .
 \label{eq:Uvertexdiag2}
 \end{eqnarray}

\vspace{7mm}

Having made the approximations $(A1)$ and $(A2)$, we are
now ready to show that {\it{in arbitrary dimension}}
the vertices  $U_{n}^{\alpha} ( q_{1} \ldots q_{n} ) $
with $n \geq 3$ vanish in the limit $q_{c} / \lambda \rightarrow 0$,
so that in this limit {\it{the Gaussian approximation becomes exact!}}
In the context of the Tomonaga-Luttinger model the vanishing of the
$U_{n}$, $n \geq 3$ has been called "weak closed loop theorem",
and is discussed
and proved in unpublished lecture notes by T. Bohr\cite{Bohr81}.
Under assumptions $(A1)$ and $(A2)$
the proof goes through in any dimension without changes.
Note that the validity of $(A1)$ and $(A2)$ is implicitly built into the
Tomonaga-Luttinger model
by definition.
The vanishing of $U_{n}$ for $n \geq 3$ is equivalent with the statement that
the
RPA for the density-density correlation function is exact in this model.
This is due to a complete cancellation between vertex- and self-energy
corrections.
In $d=1$ this cancellation has
been discovered by Dzyaloshinskii and Larkin\cite{Dzyaloshinskii74}.
Because the proof given in Ref.\cite{Dzyaloshinskii74} is rather cryptic and
the lecture notes by T. Bohr\cite{Bohr81} are not published, we briefly outline
the basic features
of the proof.
It is perhaps simpler to formulate the proof in the space-time
domain\cite{Bohr81,Hermisson95},
but for our purpose it is more convenient to work in momentum-space,
because here the Fermi-surface and the patching construction are defined.
The following two
properties of our locally linearized  Greens-function defined
in Eq.\ref{eq:Galphalin} are essential,
 \begin{equation}
 G_{0}^{\alpha} ( -  q ) =  - G_{0}^{\alpha } ( q )
 \; \; \; ,
 \label{eq:godd}
 \end{equation}
 \begin{equation}
 G_{0}^{\alpha} ( q ) G_{0}^{\alpha} ( q + q^{\prime} )
 = G^{\alpha}_{0} ( q^{\prime} )
 \left[ G_{0}^{\alpha} ( q ) - G_{0}^{\alpha} ( q + q^{\prime} )
 \right]
 \; \; \; .
 \label{eq:Gpartialfrac}
 \end{equation}
Note that Eq.\ref{eq:godd} follows trivially from the definition,
while Eq.\ref{eq:Gpartialfrac} is nothing but the partial-fraction
representation
of the product of two rational functions.
To show that the odd vertices $U_{3}, U_{5}, \ldots$ vanish, we only need
Eq.\ref{eq:godd} and the
fact that the patch $\alpha$ in Eq.\ref{eq:Uvertexdiag2} has inversion symmetry
with respect to ${\bf{k}}^{\alpha}$, so that the domain for the ${\bf{q}}$-sum
is invariant
under ${\bf{q}} \rightarrow - {\bf{q}}$.
Then it is easy to see that the contribution from a given permutation
$(P_{1} P_{2} \ldots  P_{n-1} P_{n})$ is exactly cancelled
by the contribution from the permutation
$(P_{n} P_{n-1}  \ldots P_{2} P_{1} )$ in which
the loop is traversed in the
opposite direction.
As already pointed out by T. Bohr\cite{Bohr81}, the vanishing of the odd
vertices
is a direct consequence of Furry's theorem\cite{Itzykson80}.
To show that the even vertices $U_{n}$, $n=4,6, \ldots$ vanish,
we use Eq.\ref{eq:Gpartialfrac} $n$-times for the pairs
 \begin{equation}
 \begin{array}{l}
 G_{0}^{\alpha} (q ) G_{0}^{\alpha} ( q+ q_{P_{1}} ) \; \; \; , \\
 G_{0}^{\alpha} (q +q_{P_{1}} ) G_{0}^{\alpha} ( q+ q_{P_{1}} +q_{P_{2}} )
 \; \; \; ,  \\
 \ldots \; \; \; , \\
 G_{0}^{\alpha} (q +q_{P_{1}} + \ldots + q_{P_{n-2}})
 G_{0}^{\alpha} (q +q_{P_{1}} + \ldots + q_{P_{n-1}})
 \; \; \; , \\
 G_{0}^{\alpha} (q +q_{P_{1}} + \ldots + q_{P_{n-1}}) G_{0}^{\alpha} ( q)
 \; \; \; ,
 \end{array}
 \label{eq:G0pair}
 \end{equation}
and take into account that we may replace
 $q_{P_{1}} + \ldots + q_{P_{n-1}} = - q_{P_{n}}$ because of overall
energy-momentum conservation.
Then it is easy to show that
 \begin{eqnarray}
 \lefteqn{ G_{0}^{\alpha} (q)
 G_{0}^{\alpha} ( q + q_{P_1} ) \cdots
 G_{0}^{\alpha} ( q + q_{P_1} + \ldots + q_{P_{n-1}} ) }
 \nonumber
 \\
 & = &
 \frac{1}{n}
 \left[ G_{0}^{\alpha} ( q_{P_n} ) - G_{0}^{\alpha} ( q_{P_1} ) \right]
 G_{0}^{\alpha} ( q + q_{P_1} ) \cdots
 G_{0}^{\alpha} ( q + q_{P_1} + \ldots + q_{P_{n-1}} )
 \label{eq:Uevencancel}
 \; \; \; .
 \end{eqnarray}
Substituting Eq.\ref{eq:Uevencancel} in Eq.\ref{eq:Uvertexdiag2},
shifting the summation label $q \rightarrow q -q_{P_{1}} + q_{P_{n}}$, and
using
the fact that we may rename
$q_{P_{1}} \leftrightarrow q_{P_{n}}$
because we sum
over all permutations,
it is evident that the
resulting expression vanishes.
This argument is not valid for $n=2$, because
in this case $G^{\alpha}_{0} ( q_{P_{2}} )  - G^{\alpha}_{0} ( q_{P_{1}} )
= 2
G^{\alpha}_{0} ( q_{P_{2}} ) $
due to energy-momentum conservation.
We shall discuss the vertex $U_{2}$ in detail in Sec.\ref{subsec:vertexgauss}.
Note that the shift
${{q}} \rightarrow {{q}} - {{q}}_{P_1} + {{q}}_{P_n}$ affects
also the patch-cutoff, $\Theta^{\alpha} ( {\bf{k}}^{\alpha} + {\bf{q}} )
\rightarrow
\Theta^{\alpha} ( {\bf{k}}^{\alpha} + {\bf{q}}
- {\bf{q}}_{P_1} + {\bf{q}}_{P_n} )$,
but this leads to corrections of higher order in $q_{c}$.
Because we have already ignored
higher order terms in $ q_{c} $ by making the {\it{diagonal-patch
approximation}} $(A1)$,
it is consistent to ignore this shift.

The above proof of the generalized closed loop theorem uses the language
of Greens-functions in the momentum-frequency domain.
Such a formulation naturally emerges from our functional bosonization
approach.  After completion of this work we have learnt that independently from
us W. Metzner
has also realized that the closed loop theorem
can be generalized to higher dimensions\cite{Metzner95}. His proof is based on
the analysis of the commutation relations and equations of motion
of the patch density operators $\hat{\rho}^{\alpha}_{\bf{q}}$ defined in
Eq.\ref{eq:rhoopdef}.

\vspace{7mm}

In fermionic language, the vanishing of the higher order vertices is due to
a {\it{complete cancellation between self-energy and vertex corrections}}.
This cancellation is automatically incorporated in our
bosonic formulation via the symmetrization of the vertices $U_{n}$.
We would like to emphasize again that
this remarkable cancellation happens
not only in $d=1$\cite{Dzyaloshinskii74,Bohr81}
but in arbitrary dimensions.
The existence of these cancellations in the perturbative
calculation of the dielectric function of the homogeneous
electron gas in $d=3$ has already been noticed by
Geldart and Taylor more than 20 years ago\cite{Geldart70},
although the origin for this cancellation has not been
identified. The generalized closed loop theorem
discussed in the present work gives a clear mathematical explanation
for this cancellation {\it{to all orders in perturbation theory}}.

It is important to stress that the
cancellation does not depend on the nature
of the external fields that enter the closed loop;
in particular, it occurs  also in models
where the fermionic current density is coupled
to transverse gauge-fields\cite{Kopietz95}.
The one-loop corrections to the RPA for the gauge-invariant  two-particle
Greens-functions of
electrons interacting with transverse gauge-fields
have recently calculated by Kim {\it{et al.}}\cite{Kim94}.
They found that at long wavelengths and low frequencies the
leading self-energy and vertex corrections cancel.
In the light of the generalized closed loop theorem
this cancellation is not surprising. However,
the generalized closed loop theorem is a much
stronger statement, because it implies
a cancellation between the leading self-energy and vertex corrections
to all orders
in perturbation theory.

A system where the approximations $(A1)$ and $(A2)$ are exact is
the natural generalization of the Tomonaga model to
higher dimensions\cite{Tomonaga50}.
If we follow Luttinger's construction\cite{Luttinger63}
and remove the radial cutoff $\lambda$ by extending the locally linearized
energy dispersion beyond the
squat boxes
(so that it is allowed to shift ${\bf{q}}$),
we obtain an interacting higher dimensional Fermi system that is
exactly solvable in
precisely the same sense as  the Tomonaga-Luttinger model in $d=1$.
As already mentioned, however,
for finite $q_{c} / k_{F}$  and realistic Fermi surfaces
the condition $(A1)$ can
never be exactly satisfied in $d>1$.

\section{Gaussian approximation}
\label{sec:Gauss}

In this section we shall calculate the density-density correlation
function and the bosonized Hamiltonian within the Gaussian approximation.
We also show that at long wavelengths the resulting bosonized Hamiltonian is
identical with the Hamiltonian obtained from the operator
approach\cite{Houghton94a,Castro94b}.

\subsection{The first two vertices}
\label{subsec:vertexgauss}

Within Gaussian approximation only
the vertices $U_{1}$ and $U_{2}$ are retained, and all higher order vertices
are set equal to zero.
{}From the previous subsection we know that
the Gaussian approximation is  justified
in a parameter regime where $(A1)$ and $(A2)$
are accurate.
Let us start with the vertex $U_{1}$, which is given by
 \begin{equation}
 U_{1} ( q \alpha )  =  \delta_{q,0} \frac{i}{\beta} \sum_{k}
 \Theta^{\alpha}  ( {\bf{k}} )
 \frac{1}{i \tilde{\omega}_{n} - \xi_{\bf{k}}  }
  =
  i \delta_{q,0}  N^{\alpha}_{0}
 \label{eq:U1}
 \; \; \; ,
 \end{equation}
where
 \begin{equation}
 N^{\alpha}_{0} =
 \sum_{\bf{k}}
 \Theta^{\alpha } ( {\bf{k}} )
 f (  \xi_{\bf{k}}  )
 \label{eq:N0def}
 \; \; \; ,
 \end{equation}
is the number of occupied states in
patch $\alpha$ in the non-interacting limit.
Here $f ( \epsilon ) = [ e^{\beta \epsilon } + 1 ]^{-1}$ is the Fermi-function.
Thus,
 \begin{equation}
 S_{kin,1} \left\{ \phi^{\alpha} \right\} = i  \sum_{\alpha}
  \phi^{\alpha}_{0} N^{\alpha}_{0}
 \; \; \; .
 \label{eq:Skin1}
 \end{equation}
Combining this term with the $\phi$-$\rho$-coupling in the
functional Fourier transform
in Eq.\ref{eq:Skinrhodef}, we can write
 \begin{equation}
 i \sum_{q} \sum_{\alpha} \phi^{\alpha}_{-q} \tilde{\rho}^{\alpha}_{q}
 - S_{kin,1} \left\{ \phi^{\alpha} \right\}
 =
 i \sum_{q} \sum_{\alpha} \phi^{\alpha}_{-q} \left[ \tilde{\rho}^{\alpha}_{q}
 - \delta_{q,0} N^{\alpha}_{0} \right]
 \; \; \; .
 \end{equation}
Thus,
 $S_{kin,1} \left\{ \phi^{\alpha} \right\}$ simply shifts the
collective density field according to
 \begin{equation}
 \tilde{\rho}^{\alpha}_{q} \rightarrow \tilde{\rho}^{\alpha}_{q} - \delta_{q,0}
N^{\alpha}_{0}
 \label{eq:rhoredef}
 \; \; \; .
 \end{equation}
i.e. the uniform component is shifted.
We conclude that the action $S_{kin} \left\{ \tilde{\rho}^{\alpha} \right\}$ is
actually a functional
of the shifted field. For simplicity we shall from now on
redefine the collective field according to Eq.\ref{eq:rhoredef}.
Note also that the term $q = 0$ in $S_{int} \left\{ \tilde{\rho}^{\alpha}
\right\}$
is usually excluded due to charge neutrality, so that
the effective action $S_{eff} \left\{ \tilde{\rho}^{\alpha} \right\}$
can be considered as functional of the shifted field.

\vspace{7mm}

The second-order vertex is given by
 \begin{eqnarray}
 U_{2} ( q_{1} \alpha_{1} , q_{2} \alpha_{2} )
 & =  & - \delta_{q_{1} + q_{2} , 0 }
 \frac{1}{2 \beta^2} \sum_{k}
 \left[
 \Theta^{\alpha_{1}} ( {\bf{k}} ) \Theta^{\alpha_{2}} ( {\bf{k}}+ {\bf{q}}_{1}
)
 G_{0} ( k ) G_{0} ( k + q_{1} )
 \right.
 \nonumber
 \\
 & & +
 \left.
 \Theta^{\alpha_{2}} ( {\bf{k}} ) \Theta^{\alpha_{1}} ( {\bf{k}}+ {\bf{q}}_{2}
)
 G_{0} ( k ) G_{0} (  k + q_{2} )
 \right]
 \label{eq:U2def}
 \; \; \; .
 \end{eqnarray}
Performing the frequency sum
we obtain
 \begin{equation}
 U_{2}( -q \alpha ,q \alpha^{\prime} ) =
\frac{V}{\beta}
\Pi_{0}^{\alpha \alpha^{\prime} } (q)
\label{eq:U2res}
\; \; \; ,
\end{equation}
where
 \begin{equation}
 \Pi_{0}^{\alpha \alpha^{\prime} }
 ( q )
    =   \frac{1}{2 {V}} \sum_{\bf{k}}
  \left[
 \Theta^{\alpha} ( {\bf{k}} ) \Theta^{\alpha^{\prime} }
 ( {\bf{k-q}} )
\frac{ f (  \xi_{\bf{k-q}}  )
 - f (  \xi_{\bf{k}} ) }
{ \xi_{\bf{k}} - \xi_{\bf{k-q}} - i \omega_{m} }
  +
 \left( \alpha \leftrightarrow \alpha^{\prime} , q \rightarrow -q \right)
 \right]
 \label{eq:Pi0}
 \; \; \; .
 \end{equation}
Hence $S_{kin,2} \{ \phi^{\alpha} \}$ is given by
 \begin{equation}
 S_{kin,2} \{ \phi^{\alpha} \} =
 \frac{V}{2 \beta } \sum_{q} \sum_{\alpha \alpha^{\prime}}
 \Pi_{0}^{\alpha \alpha^{\prime}} (q)
 \phi^{\alpha}_{-q} \phi^{\alpha^{\prime}}_{q}
 \label{eq:Skin2phires}
 \; \; \; .
 \end{equation}
Note that $\Pi_{0}^{\alpha \alpha^{\prime} }(q)$ is the patch polarization
in the absence of interactions, i.e. the exact result
of Eq.\ref{eq:cordensloc} if we set $\underline{\tilde{f}}_{q} = 0$.
In this case our complicated transformations
are of course not necessary and it is
much simpler to calculate the density-density correlation function
directly from its definition in terms of fermionic operators.
For $|{\bf{q}}| \ll k_{F}$ the {\it{diagonal-patch approximation}}
$(A1)$ is justified, so that
in Eq.\ref{eq:Pi0} we may replace
$ \Theta^{\alpha} ( {\bf{k}} ) \Theta^{ \alpha^{\prime} }
( {\bf{k-q}} )  \approx
\delta^{\alpha \alpha^{\prime}}
\Theta^{\alpha} ( {\bf{k}} )  $.
To leading order in $|{\bf{q}} | / k_{F}$ we have therefore in any dimension
 \begin{equation}
 \Pi^{\alpha \alpha^{\prime} }_{0} ( q ) \approx
 \delta^{\alpha \alpha^{\prime} } \Pi^{\alpha}_{0} ( q )
 \; \; \; , \; \; \;
 \Pi^{\alpha}_{0} ( q) =
 \nu^{\alpha}
 \frac{  {\bf{v}}^{\alpha} \cdot {\bf{q}} }
 {  {\bf{v}}^{\alpha} \cdot {\bf{q}} - i \omega_{m} }
 \; \; \; ,
 \label{eq:Pilong}
 \end{equation}
where
 \begin{equation}
 \nu^{\alpha} = \frac{1}{V} \frac{ \partial N^{\alpha}_{0} }{\partial \mu }
 =
  \frac{1}{V} \sum_{\bf{k}} \Theta^{\alpha} ( {\bf{k}} )
  \left[ - \frac{ \partial f ( \xi_{\bf{k}} ) }{\partial \xi_{\bf{k}} } \right]
 \label{eq:nualphadef}
 \end{equation}
is the "local" density of states associated with patch $\alpha$, and
${\bf{v}}^{\alpha}$ is the local Fermi velocity, see Eq.\ref{eq:valphadef}.
Note that Eq.\ref{eq:Pilong} is
valid  for small $|{\bf{q}}| / k_{F}$ but for {\it{arbitrary}} frequencies.

\subsection{Density-density correlation function}

Within Gaussian approximation the
action for the dual field $\phi^{\alpha}_{q}$ is quadratic, so that the
functional Fourier transform in Eq.\ref{eq:Skinrhodef}
can be calculated trivially, with the result
 \begin{equation}
  S_{kin} \left\{ \tilde{\rho}^{\alpha} \right\} =
 S_{kin,0}^{(0)} +
  \frac{1}{2} \sum_{q} \sum_{\alpha \alpha^{\prime}}
  \Gamma^{\alpha \alpha^{\prime}}_{q} \tilde{\rho}^{\alpha}_{-q}
\tilde{\rho}^{\alpha^{\prime}}_{q}
  \label{eq:Skingaus}
  \; \; \; ,
  \end{equation}
where $S_{kin,0}^{(0)}$ is a constant independent of the
$\tilde{\rho}^{\alpha}$-field, and
$\Gamma_{q}^{\alpha \alpha^{\prime} }$ is the propagator of the
$\phi^{\alpha}$-field
with respect to the Gaussian action $S_{kin,2} \left\{ \phi^{\alpha} \right\}$,
 \begin{equation}
 \left< \phi^{\alpha}_{q} \phi^{\alpha^{\prime} }_{q^{\prime}} \right>_{0}
 = \delta_{q + q^{\prime} , 0} \Gamma_{q}^{\alpha \alpha^{\prime}}
 \label{eq:gammaprop}
 \; \; \; .
 \end{equation}
Here $< \ldots >_{0}$ denotes Gaussian average
with probability measure $e^{-S_{kin,2} \{ \phi^{\alpha} \} }$, i.e.
for any functional $F \left\{ \phi^{\alpha} \right\}$
 \begin{equation}
 \left< F \left\{ \phi^{\alpha} \right\} \right>_{0 }
  =
 \frac{
 \int {\cal{D}} \left\{ \phi^{\alpha} \right\}
  F \left\{ \phi^{\alpha} \right\}
 e^{ - S_{kin,2} \left\{ \phi^{\alpha} \right\}  }
  }
 {
 \int {\cal{D}} \left\{ \phi^{\alpha} \right\}
 e^{
  - S_{kin,2} \left\{ \phi^{\alpha} \right\}  }
  }
 \label{eq:funcrho3}
 \; \; \; .
 \end{equation}
Note that $\Gamma_{q}^{\alpha \alpha^{\prime}}$ is
the inverse of
the matrix $U_{2} ( -q \alpha , q \alpha^{\prime} )$ in the space spanned by
the
patch-indices, i.e.
 \begin{equation}
 \sum_{\alpha^{\prime}} U_{2} ( -q \alpha , q \alpha^{\prime} )
\Gamma_{q}^{\alpha^{\prime} \alpha^{\prime \prime}}
 = \delta^{ \alpha \alpha^{\prime \prime} }
 \; \; \; .
 \label{eq:Gammapropdef}
 \end{equation}
Thus, $\Gamma_{q}^{\alpha \alpha^{\prime}}$ is proportional to the
{\it{inverse}} of the non-interacting patch polarization.
Absorbing for simplicity the constant $S_{kin,0}^{(0)}$ in the re-definition
of the integration measure ${\cal{D}} \left\{ \tilde{\rho}^{\alpha} \right\}$
and combining
the contribution from the kinetic energy with the
interaction contribution, we obtain for
the bosonized action $S_{eff} \left\{ \tilde{\rho}^{\alpha} \right\}$
defined in Eq.\ref{eq:Seffrhodef} in Gaussian approximation
 \begin{equation}
 {S}_{eff} \left\{ \tilde{\rho}^{\alpha} \right\}   =
 \frac{1}{2} \sum_{q} \sum_{\alpha \alpha^{\prime}}
 \left[ [ \underline{\tilde{f}}_{q} ]^{ \alpha \alpha^{\prime} }
  + \Gamma_{q}^{\alpha \alpha^{\prime}} \right]
 \tilde{\rho}^{\alpha}_{-q} \tilde{\rho}^{\alpha^{\prime}}_{q}
 \label{eq:Seffgauss}
 \; \; \; .
 \end{equation}
Note that the inclusion of interactions is trivial, while the major difficulty
lies in the
representation of the kinetic energy as a functional of the
collective density field.
The Gaussian propagator of the $\tilde{\rho}^{\alpha}$-field is then
 \begin{equation}
 \frac{
 \int
 {\cal{D}} \left\{ \tilde{\rho}^{\alpha} \right\}
 e^{- {S}_{eff} \left\{ \tilde{\rho}^{\alpha} \right\} }
 \tilde{\rho}^{\alpha}_{q} \tilde{\rho}^{\alpha^{\prime}}_{q^{\prime}}
}
 {
 {\cal{D}} \left\{ \tilde{\rho}^{\alpha} \right\}
 e^{- {S}_{eff} \left\{ \tilde{\rho}^{\alpha} \right\} }
 }
 = \delta_{q + q^{\prime} , 0}
 \left[ \left[  \underline{\tilde{f}}_{q} + \underline{\Gamma}_{q} \right]^{-1}
\right]^{\alpha \alpha^{\prime}}
 \label{eq:rhopropgauss}
 \; \; \; ,
 \end{equation}
where $\underline{\Gamma}_{q}$ is a matrix in the patch-labels, with
$[ \underline{\Gamma}_{q} ]^{\alpha \alpha^{\prime} } = \Gamma_{q}^{\alpha
\alpha^{\prime}}$.
{}From Eq.\ref{eq:Pirho} we obtain for the density-density correlation function
 \begin{eqnarray}
 \Pi^{\alpha  \alpha^{\prime} } ( q )
  & = & \frac{ \beta }{V}
 \left[ \left[   \underline{\tilde{f}}_{q} + \underline{\Gamma}_{q}
\right]^{-1} \right]^{\alpha \alpha^{\prime}}
 \nonumber
 \\
 & = &
 \left[ \left[
 \underline{\Pi}_{0}^{-1} ( q )  +
 \underline{f}_{q} \right]^{-1}
 \right]^{\alpha \alpha^{\prime}}
 \label{eq:densrpa}
 \; \; \; ,
 \end{eqnarray}
where $\underline{\Pi}_{0} (q )$ is again a matrix in the patch-labels, with
 \begin{equation}
 [ \underline{\Pi}_{0} ( q )]^{\alpha \alpha^{\prime}} =
 \frac{\beta}{V}
 \left[ \underline{\Gamma}^{-1}_{q} \right]^{\alpha \alpha^{\prime}}
 \equiv
 \Pi_{0}^{\alpha \alpha^{\prime}} ( q)
 \label{eq:Pihat0}
 \; \; \; ,
 \end{equation}
see Eqs.\ref{eq:U2res}, \ref{eq:Pi0} and \ref{eq:Gammapropdef}.
The matrix elements of $\underline{f}_{q}$ are the usual Landau-parameters
defined in
Eq.\ref{eq:Landaufuncdef}, i.e.
 $[ \underline{{f}}_{q} ]^{\alpha \alpha^{\prime} } =
 f_{\bf{q}}^{\alpha \alpha^{\prime}} $.
Eq.\ref{eq:densrpa} is nothing but the RPA
for the {\it{patch}} density-density correlation function.
To obtain the standard RPA-result for the total density-density
correlation function, we should
according to Eq.\ref{eq:Ptotdecompose}
sum Eq.\ref{eq:densrpa} over both patch-labels,
 \begin{equation}
 \Pi ( q) =
 \sum_{\alpha \alpha^{\prime}}
 \left[
 \left[ \underline{\Pi}_{0}^{-1} ( q ) + \underline{f}_{q}  \right]^{-1}
 \right]^{\alpha \alpha^{\prime}}
 \; \; \; .
 \label{eq:Ptotdecompose1}
 \end{equation}
For simplicity let us assume that $[ \underline{f}_{q} ]^{\alpha
\alpha^{\prime} } = {f}_{\bf{q}}$ is
independent of the patch-indices.
Expanding
 \begin{equation}
  [ \underline{\Pi}_{0}^{-1} (q) + \underline{f}_{q} ]^{-1}
 = \underline{\Pi}_{0} (q)  - \underline{ \Pi}_{0} (q) \underline{f}_{{q}}
\underline{\Pi}_{0} ( q) +
 \underline{ \Pi}_{0} (q) \underline{f}_{{q}} \underline{\Pi}_{0} ( q)
 \underline{f}_{{q}} \underline{\Pi}_{0} ( q)
  - \ldots
  \label{eq:Neumann}
  \; \; \; ,
  \end{equation}
and taking matrix elements, we see that Eqs.\ref{eq:Ptotdecompose1} and
\ref{eq:Neumann}
reduce to
 \begin{equation}
 \Pi ( q) = \frac{ \Pi_{0} (q) }{1 + f_{\bf{q}} \Pi_{0} (q) }
 \label{eq:Pi0tot}
 \; \; \; ,
 \end{equation}
where
 \begin{equation}
 \Pi_{0} (q)
 = \sum_{\alpha \alpha^{\prime}} \Pi_{0}^{\alpha \alpha^{\prime}} (q)
 \label{eq:pi0tot}
 \end{equation}
is the {\it{total}} non-interacting polarization.
We would like to emphasize that up to this point we have not linearized the
energy dispersion, so that Eq.\ref{eq:Pi0tot} is the exact RPA result
for all wave-vectors,
including the short wavelength regime.

\subsection{Bosonization of the Hamiltonian}

To make contact with the operator-formalism to
bosonization\cite{Houghton93}-\cite{Castro94c}
let us now derive a bosonic Hamiltonian
that is in the limit of long wavelengths
equivalent to our Gaussian action $S_{eff} \left\{ \tilde{\rho}^{\alpha}
\right\}$
in Eq.\ref{eq:Seffgauss}.
The key observation is that the patch
polarization is
{\it{in the limit of high densities and long wavelengths}}
(i.e. in the limit where the {\it{diagonal-patch approximation}} $(A1)$ is
correct)
diagonal in the patch indices, and of the form given in Eq.\ref{eq:Pilong}.
{}From Eqs.\ref{eq:U2res}, \ref{eq:Pi0} and \ref{eq:Gammapropdef} it is then
easy to see that
the matrix elements of $\underline{\Gamma}_{q}$ are given by
 \begin{equation}
 \Gamma_{q}^{\alpha \alpha^{\prime} } \approx \delta^{ \alpha \alpha^{\prime}}
 \frac{ \beta}{V \nu^{\alpha}}
 \frac{  {\bf{v}}^{\alpha} \cdot {\bf{q}} - i \omega_{m} }
 {  {\bf{v}}^{\alpha} \cdot {\bf{q}} }
 \label{eq:Gammalong}
 \; \; \; , \; \; \;  | {\bf{q}} | \ll k_{F}
 \; \; \; .
 \end{equation}
Hence the Gaussian action in Eq.\ref{eq:Seffgauss} can be written as
\begin{equation}
 {S}_{eff} \left\{ \tilde{\rho}^{\alpha} \right\}   =
 \frac{\beta}{2V} \sum_{q} \sum_{\alpha \alpha^{\prime}}
 \left[ {f}_{q}^{ \alpha \alpha^{\prime} }
  +
  \delta^{\alpha \alpha^{\prime}}
 \frac{  {\bf{v}}^{\alpha} \cdot {\bf{q}} - i \omega_{m} }
 { \nu^{\alpha}  {\bf{v}}^{\alpha} \cdot {\bf{q}} }
  \right]
 \tilde{\rho}^{\alpha}_{-q} \tilde{\rho}^{\alpha^{\prime}}_{q}
 \label{eq:Seffgauss2}
 \; \; \; .
 \end{equation}
Obviously the term proportional to $i \omega_{m}$
defines the dynamics of the $\tilde{\rho}^{\alpha}$-field.
We now recall that in the
functional integral for canonically quantized bosons
{\it{the coefficient of the term proportional to $-i \omega_{m}$ should be
precisely
$\beta$.}}  Thus, to write our effective action in terms of a canonical
boson field $b_{q}^{\alpha}$, we should rescale the
$\tilde{\rho}^{\alpha}$-field
accordingly.
This is achieved by substituting
 \begin{equation}
 \tilde{\rho}^{\alpha}_{q} = ({ {{V}} \nu^{\alpha} |
 {\bf{v}}^{\alpha} \cdot {\bf{q}} | })^{\frac{1}{2}}
 \left[
 \Theta ( {\bf{v}}^{\alpha} \cdot {\bf{q}} ) b^{ \alpha}_{q} +
 \Theta ( - {\bf{v}}^{\alpha} \cdot {\bf{q}} ) b^{\dagger \alpha}_{-q}
 \right]
 \label{eq:rhobscale}
 \end{equation}
in Eq.\ref{eq:Seffgauss2}.
The $\Theta$-functions are necessary to make the coefficient
of $-i \omega_{m}$ equal to $\beta$ for all patches, because the sign of $i
\omega_{m}$ in
Eq.\ref{eq:Seffgauss2} depends on the sign of ${ {\bf{v}}^{\alpha} \cdot
{\bf{q}}}$.
Our final result for the
bosonized action
$ S \left\{ b^{\alpha} \right\} \equiv S_{eff} \left\{ \tilde{\rho}^{\alpha}
(b^{\alpha}) \right\}$
is
 \begin{eqnarray}
 S \left\{ b^{\alpha} \right\}
 & = &
  \beta \sum_{q} \sum_{\alpha  } \Theta ( {\bf{v}}^{\alpha } \cdot {\bf{q}} )
  ( - i \omega_{m} )
 b^{ \alpha \dagger }_{q} b^{\alpha}_{q}
 \nonumber
 \\
 & + &
  \beta  \left[ H_{kin}
 \left\{ b^{\alpha} \right\}
 +  H_{int}
 \left\{ b^{\alpha} \right\} \right]
 \; \; \; ,
 \label{eq:Sb2}
 \end{eqnarray}
 \begin{eqnarray}
 H_{kin} \left\{ b^{\alpha} \right\} & = &
   \sum_{q} \sum_{\alpha  } \Theta ( {\bf{v}}^{\alpha } \cdot {\bf{q}} )
  {\bf{v}}^{\alpha} \cdot {\bf{q}}
 b^{\alpha \dagger }_{q} b^{\alpha}_{q}
 \; \; \; ,
 \label{eq:Hkin}
 \\
 H_{int} \left\{ b^{\alpha} \right\} & = &
 \frac{1}{2} \sum_{q} \sum_{\alpha \alpha^{\prime} }
 \Theta ( {\bf{v}}^{\alpha } \cdot {\bf{q}} )
  \sqrt{ |  {\bf{v}}^{\alpha} \cdot {\bf{q}} |
 |  {\bf{v}}^{\alpha^{\prime}} \cdot {\bf{q}} |}
 \nonumber
 \\
 &  & \hspace{-11mm} \times \left[
 \Theta ( {\bf{v}}^{\alpha^{\prime} } \cdot {\bf{q}} )
 \left(
 \bar{f}_{\bf{q}}^{\alpha \alpha^{\prime} }
 b^{\alpha \dagger }_{q} b^{\alpha^{\prime}}_{q}
 +
 \bar{f}_{\bf{q}}^{\alpha^{\prime} \alpha }
 b^{ \alpha^{\prime} \dagger }_{q} b^{\alpha}_{q}
 \right)
 \right.
 \nonumber
 \\
 &  &  \hspace{-10mm} +
 \left.
 \Theta ( - {\bf{v}}^{\alpha^{\prime} } \cdot {\bf{q}} )
 \left(
 \bar{f}_{\bf{q}}^{\alpha \alpha^{\prime} }
 b^{ \alpha \dagger }_{q} b^{ \alpha^{\prime} \dagger }_{-q}
 +
 \bar{f}_{\bf{q}}^{\alpha^{\prime} \alpha }
 b^{\alpha^{\prime}}_{-q} b^{\alpha}_{q}
 \right)
 \right]
 \label{eq:Hint}
 \;  \; ,
 \end{eqnarray}
where $\bar{f}_{\bf{q}}^{\alpha \alpha^{\prime} } =
 \sqrt{ \nu^{\alpha} \nu^{\alpha^{\prime}} } f^{\alpha
\alpha^{\prime}}_{\bf{q}}$ are
dimensionless couplings.
The functional integral for the $b^{\alpha}$-field
is now formally identical with a
standard bosonic functional integral.
The corresponding second quantized bosonic Hamiltonian
is therefore $\hat{H}^{b} = \hat{H}^{b}_{kin} + \hat{H}^{b}_{int}$,
where $\hat{H}^{b}_{kin}$ and $\hat{H}^{b}_{int}$ are simply
obtained by replacing
the bosonic fields $b^{\alpha}_{q}$
in Eqs.\ref{eq:Hkin},\ref{eq:Hint} by operators
$\hat{b}^{\alpha}_{\bf{q}}$ satisfying
$[ \hat{b}^{\alpha}_{\bf{q}} , \hat{b}^{\alpha^{\prime} \dagger }_{
{\bf{q}}^{\prime} }
 ] = \delta^{ \alpha \alpha^{\prime} } \delta_{ {\bf{q}} , {\bf{q}}^{\prime}
}$.
The resulting $\hat{H}^{b}$ agrees with
the bosonized Hamiltonian derived
in Refs.\cite{Houghton94a,Castro94b}
by means of an operator approach.

Note, however,
that the above identification with a canonical bosonic Hamiltonian is only
possible in the limit of long wavelengths and high densities,
so that our parametrization of the effective Gaussian action
in Eq.\ref{eq:Seffgauss}
is more general.
Moreover, for practical calculations the
substitution in Eq.\ref{eq:rhobscale} is not very useful, because
it maps the very simple form of $S_{eff} \left\{ \tilde{\rho}^{\alpha}
\right\}$
given in Eq.\ref{eq:Seffgauss} onto the complicated effective
action $S \left\{ b^{\alpha} \right\}$ in Eqs.\ref{eq:Sb2}-\ref{eq:Hint}
without
containing new information.

\section{Beyond the Gaussian approximation}
\setcounter{equation}{0}
\label{sec:beyond}

In this section we shall  calculate the one-loop correction to the
Gaussian approximation for the effective action of the collective density
field.
In this way we determine the hidden parameter which
determines  the range of validity of the Gaussian approximation.
We also show that bosonization leads to a new systematic method to calculate
corrections
to the RPA for the dielectric function.

\subsection{General perturbative expansion of $S_{kin} \left\{
\tilde{\rho}^{\alpha} \right\}$ }

In this subsection we develop a systematic perturbative method
to calculate the bosonized kinetic energy
$S_{kin} \left\{ \tilde{\rho}^{\alpha} \right\}$ defined in
Eq.\ref{eq:Skinrhodef}.
This method is based on a loop-wise expansion of the
functional Fourier transformation with the help of the
linked cluster theorem.
Defining $S_{kin}^{\prime} \left\{ \phi^{\alpha} \right\}$ to be the sum
of all non-Gaussian terms in the expansion of the effective
action of the dual field (see Eq.\ref{eq:Skinexp}),
 \begin{equation}
 S_{kin}^{\prime} \left\{ \phi^{\alpha} \right\} = \sum_{n=3}^{\infty}
S_{kin,n}
 \left\{ \phi^{\alpha} \right\}
 \label{eq:Skin3}
 \; \; \; ,
 \end{equation}
we need to calculate
 \begin{equation}
 e^{- S_{kin} \left\{ \tilde{\rho}^{\alpha} \right\} }
  = e^{- S_{kin,0}^{(0)} -
  S_{kin,2}^{(0)} \left\{ \tilde{\rho}^{\alpha} \right\} }
 \left< e^{ -
 S_{kin}^{\prime} \left\{ \phi^{\alpha} \right\}  } \right>_{\tilde{\rho} }
 \label{eq:Skinrho2}
 \; \; \; ,
 \end{equation}
where we have defined
 \begin{equation}
  S_{kin,2}^{(0)} \left\{ \tilde{\rho}^{\alpha} \right\}
  =
  \frac{1}{2} \sum_{q} \sum_{\alpha \alpha^{\prime}}
  \Gamma^{\alpha \alpha^{\prime}}_{q} \tilde{\rho}^{\alpha}_{-q}
\tilde{\rho}^{\alpha^{\prime}}_{q}
  \label{eq:Skin20}
  \; \; \; ,
  \end{equation}
and
 \begin{equation}
 \left< F \left\{ \phi^{\alpha} \right\} \right>_{\tilde{\rho} }
 =
 \frac{
 \int {\cal{D}} \left\{ \phi^{\alpha} \right\}
  F \left\{ \phi^{\alpha} \right\}
 e^{
  i   \sum_{q \alpha} \phi^{\alpha}_{-q} \tilde{\rho}^{\alpha}_{q}
  - S_{kin,2} \left\{ \phi^{\alpha} \right\}  }
  }
 {
 \int {\cal{D}} \left\{ \phi^{\alpha} \right\}
 e^{
  i   \sum_{q \alpha} \phi^{\alpha}_{-q} \tilde{\rho}^{\alpha}_{q}
  - S_{kin,2} \left\{ \phi^{\alpha} \right\}  }
  }
 \label{eq:funcrhodef}
 \end{equation}
for any functional $F \left\{ \phi^{\alpha} \right\}$.
Performing the shift-transformation
 \begin{equation}
 \phi^{\alpha}_{q} \rightarrow {\phi}^{\alpha}_{q} +
 i \sum_{\alpha^{\prime}}
 \Gamma_{q}^{\alpha \alpha^{\prime} } \tilde{\rho}_{q}^{\alpha^{\prime}}
 \label{eq:shift1}
 \; \; \; ,
 \end{equation}
it is easy to see that
 \begin{eqnarray}
 \left< F \left\{ \phi^{\alpha} \right\} \right>_{\tilde{\rho} }
  =
 \left< F
 \{ \phi^{\alpha} + i
 \sum_{\alpha^{\prime}}
 \Gamma^{\alpha \alpha^{\prime} } \tilde{\rho}^{\alpha^{\prime}}
 \}
 \right>_{0 }
 \label{eq:funcrho2}
 \; \; \; .
 \end{eqnarray}
In our case we have to calculate
 \begin{equation}
 \left< e^{ -
 S_{kin}^{\prime} \left\{ \phi^{\alpha} \right\}  } \right>_{\tilde{\rho} }
 =
 \left< e^{ -
 S_{kin}^{\prime} \{ \phi^{\alpha}
 + i \sum_{\alpha^{\prime}}
 \Gamma^{\alpha \alpha^{\prime}
 } \tilde{\rho}^{\alpha^{\prime}} \}
  } \right>_{0 }
 \label{eq:Skinexprho}
 \; \; \; .
 \end{equation}
Consider now the term of order $(\phi^{\alpha} )^n$ in the expansion
of $S^{\prime}_{kin} \left\{ \phi^{\alpha} \right\}$
given in Eq.\ref{eq:Skin3}.
Clearly the substitution
$ \phi^{\alpha}_{q} \rightarrow
 \phi^{\alpha}_{q}
 + i \sum_{\alpha^{\prime}}
 \Gamma^{\alpha \alpha^{\prime}
 }_{q} \tilde{\rho}^{\alpha^{\prime}}_{q}$ generates also a term of order
$(\tilde{\rho}^{\alpha})^n$
which does not depend on the
$\phi^{\alpha}$-field, and can be pulled out of the average in
Eq.\ref{eq:Skinexprho}.
Let us denote this term by
$S_{kin,n}^{(0)} \{ \tilde{\rho}^{\alpha} \}$.
{}From Eq.\ref{eq:Seffphin} it is easy to see that
$S_{kin,n}^{ (0)} \{ \tilde{\rho}^{\alpha} \}$
is obtained by replacing
$ \phi^{\alpha}_{q} \rightarrow
  i \sum_{\alpha^{\prime}}
 \Gamma^{\alpha \alpha^{\prime}
 }_{q} \tilde{\rho}^{\alpha^{\prime}}_{q}$
in $S_{kin,n} \{ \phi^{\alpha} \}$, so that
it is given by
 \begin{eqnarray}
 S_{kin,n}^{(0)}
 \left\{ { { \tilde{\rho}^{\alpha}}} \right\} & = &
S_{kin,n} \{
  i \sum_{\alpha^{\prime}}
 \Gamma^{\alpha \alpha^{\prime}
 }_{q} \tilde{\rho}^{\alpha^{\prime}}_{q}
\} =
\nonumber
\\
&  & \hspace{-19mm}
 \frac{1}{n} \sum_{q_{1}  \ldots q_{n} }
 \sum_{\alpha_{1} \ldots \alpha_{n} }
 \Gamma_{n}^{(0)} (
 q_1 \alpha_{1} \ldots q_{n} \alpha_{n}  ) \tilde{\rho}^{\alpha_{1}}_{q_{1}}
\cdots
 \tilde{\rho}^{\alpha_{n}}_{q_{n}}
 \; \; \; ,
 \label{eq:Seffrhon0}
 \end{eqnarray}
where for $n \geq 3$ the vertices $\Gamma_{n}^{(0)}$ are
 \begin{equation}
 \Gamma_{n}^{(0)} (
 q_1 \alpha_{1} \ldots q_{n} \alpha_{n}  )
 = i^{n} \sum_{\alpha_{1}^{\prime} \ldots \alpha_{n}^{\prime} }
 U_{n} (
 q_1 \alpha_{1}^{\prime} \ldots q_{n} \alpha_{n}^{\prime}  )
 \Gamma_{q_{1}}^{\alpha_{1}^{\prime} \alpha_{1}} \ldots
 \Gamma_{q_{n}}^{\alpha_{n}^{\prime} \alpha_{n}}
 \label{eq:Gammandef0}
 \; \; \; , \; \; \; n \geq 3 \; \; \; .
 \end{equation}
Obviously the Gaussian action $S^{(0)}_{kin,2} \{ \rho^{\alpha} \}$
in Eq.\ref{eq:Skin20} is also of the form \ref{eq:Seffrhon0}, with
 \begin{equation}
 \Gamma_{2}^{(0)} ( q_{1} \alpha_{1} q_{2} \alpha_{2} )
 = \delta_{q_{1} + q_{2} , 0} \Gamma^{\alpha_{1} \alpha_{2} }_{q_{2}}
 \; \; \; .
 \label{eq:Gamma20Gauss}
 \end{equation}
The vertex $U_{1}$ has been absorbed into the re-definition
of $\tilde{\rho}^{\alpha}_{q}$ (see Eq.\ref{eq:rhoredef}), so that
$S_{kin,1}^{(0)} \{ \tilde{\rho}^{\alpha}_{q} \} = 0$.
Defining
 \begin{eqnarray}
 S_{kin}^{(0)}
 \left\{ { { \tilde{\rho}^{\alpha}}} \right\}  &   = &
 S_{kin,0}^{(0)} +
 \sum_{n=2}^{\infty} S_{kin,n}^{(0)}
 \left\{ { { \tilde{\rho}^{\alpha}}} \right\}
 \; \; \; ,
 \label{eq:Skin0rho}
 \\
 S^{\prime \prime}_{kin} \left\{ \phi^{\alpha} , \tilde{\rho}^{\alpha} \right\}
 & = &
 S_{kin}^{\prime} \{ \phi^{\alpha}
 + i \sum_{\alpha^{\prime}}
 \Gamma^{\alpha \alpha^{\prime}
 } \tilde{\rho}^{\alpha^{\prime}} \}
 -
S_{kin}^{\prime} \{
  i \sum_{\alpha^{\prime}}
 \Gamma^{\alpha \alpha^{\prime} } \tilde{\rho}^{\alpha^{\prime}}
\}
 \label{eq:Sprimeprime}
 \; \; \; ,
 \end{eqnarray}
the general perturbative expansion for $S_{kin} \left\{ \tilde{\rho}^{\alpha}
\right\}$ is
 \begin{equation}
S_{kin} \left\{ \tilde{\rho}^{\alpha} \right\}
= S^{(0)}_{kin}
\left\{ \tilde{\rho}^{\alpha} \right\}
- \ln \left[ 1 +
 \sum_{k=1}^{\infty} \frac{(-1)^{k}}{k ! }
 \left< \left[
 S_{kin}^{\prime \prime} \{ \phi^{\alpha} , \tilde{\rho}^{\alpha} \}
 \right]^k \right>_{0}
 \right]
 \label{eq:generalpert}
 \; \; \; .
 \end{equation}
According to the linked cluster theorem\cite{Mahan81} the
logarithm eliminates all disconnected diagrams, so that Eq.\ref{eq:generalpert}
can also be written as
 \begin{equation}
S_{kin} \left\{ \tilde{\rho}^{\alpha} \right\}
= S^{(0)}_{kin}
\left\{ \tilde{\rho}^{\alpha} \right\}
-
 \sum_{k=1}^{\infty} \frac{(-1)^{k}}{k  }
 \left<
 \left[
 S_{kin}^{\prime \prime} \{ \phi^{\alpha} , \tilde{\rho}^{\alpha} \}
 \right]^k \right>_{0,c}
 \label{eq:generalpert1}
 \; \; \; ,
 \end{equation}
where the subscript $c$ means that all different connected diagrams should
be retained\cite{Mahan81}.
{}From this expression it is easy to see that
$S_{kin} \left\{ \tilde{\rho}^{\alpha} \right\}$ is in general of the following
form
 \begin{equation}
 S_{kin} \left\{ \tilde{\rho}^{\alpha} \right\}
 =  S_{kin,0} +  \sum_{n=1}^{\infty} S_{kin , n}
 \left\{ \tilde{\rho}^{\alpha} \right\}
 \label{eq:Skinn}
 \; \; \; ,
 \end{equation}
where $S_{kin,0}$ is a constant
independent of the fields that cancels in the calculation of correlation
functions, and
for $n \geq 1$
 \begin{equation}
 S_{kin,n} \left\{ { { \tilde{\rho}^{\alpha}}} \right\}  =
 \frac{1}{n} \sum_{q_{1}  \ldots q_{n} }
 \sum_{\alpha_{1} \ldots \alpha_{n} }
 \Gamma_{n} (
 q_1 \alpha_{1} \ldots q_{n} \alpha_{n}  ) \tilde{\rho}^{\alpha_{1}}_{q_{1}}
\cdots
 \tilde{\rho}^{\alpha_{n}}_{q_{n}}
 \; \; \; ,
 \label{eq:Seffrhon}
 \end{equation}
where the vertices $\Gamma_{n}$ have an expansion of the form
 \begin{equation}
 \Gamma_{n} (
 q_1 \alpha_{1} \ldots q_{n} \alpha_{n}  )
 = \sum_{k = 0}^{\infty}
 \Gamma_{n}^{(k)} (
 q_1 \alpha_{1} \ldots q_{n} \alpha_{n}  )
 \; \; \; .
 \label{eq:Gammank}
 \end{equation}
Here $\Gamma_{n}^{(k)}$
describes the interaction between $n$ collective density fields
$\tilde{\rho}^{\alpha}_{q}$,
that are generated  from all diagrams in the linked cluster expansion
\ref{eq:generalpert1}
containing $k$ internal loops of the $\phi^{\alpha}$-field.
Note that the vertices $\Gamma_{n}^{(0)}$ in Eq.\ref{eq:Gammandef0}
do not contain any internal $\phi^{\alpha}$-loops, and are therefore the
tree-approximation for the exact vertices $\Gamma_{n}$.
Because each internal $\phi^{\alpha}$-loop attached to a vertex $U_{m}$ reduces
the number
of external $\phi$-fields by $2$, it is clear that
for $k \geq 1$ the vertices
$\Gamma_{n}^{(k)}$ can only be determined by vertices $U_{m}$ with $m > n$.
Within the Gaussian approximation the vertices $U_{m}$ with $m \geq 3$ are set
equal to zero,
while the contribution from $U_{1}$ can be absorbed into the re-definition
of the $q=0$-part of $\tilde{\rho}^{\alpha}_{q}$, see Eq.\ref{eq:rhoredef}.
Hence in Gaussian approximation
 \begin{equation}
 \Gamma_{2} ( -q \alpha , q \alpha^{\prime} ) \approx
 \Gamma^{(0)}_{2} ( -q \alpha , q \alpha^{\prime} ) = \Gamma_{q}^{\alpha
\alpha^{\prime} }
 \label{eq:gaussgamma1}
 \; \; \; ,
 \end{equation}
where $\Gamma_{q}^{\alpha \alpha^{\prime} }$ is defined in
Eq.\ref{eq:Gammapropdef},
and
 \begin{equation}
 \Gamma_{n}^{(k)} = 0 \; \; \; , \; \; \; \mbox{for $n \geq 3$ or $k \geq 1$}
 \label{eq:gaussgamma2}
 \; \; \; .
 \end{equation}
Although $\Gamma_{1} = 0$ at the level of the Gaussian approximation, the
higher
order terms will in general lead to a finite value of
$\Gamma_{1}$, which describes the fluctuations of the total number
of occupied states in the patches. These terms do not contribute to
correlation functions at finite $q$, but are certainly important
for the calculation of the free energy.
The formalism developed above is the starting point for a systematic
calculation of corrections to the Gaussian approximation.

\subsection{Explicit calculation of the leading correction to the Gaussian
approximation}

The leading correction to the Gaussian approximation is obtained from the
one-loop correction in our
effective
bosonic theory, which amounts to a two-loop calculation at the fermionic level.
Note that we have mapped the problem of calculating
a two-particle Greens-function of the original fermionic model onto the
problem of calculating a one-particle Greens-function of an effective
bosonic model. The latter is conceptually  simpler, because the
symmetrized vertices $U_{n}$ and $\Gamma_{n}^{(k)}$ automatically contain
the relevant self-energy and vertex corrections of the underlying
fermionic problem.
This will become evident below.

At one-loop order, it is sufficient to truncate the
expansion of the interaction part $S^{\prime}_{kin} \{ \phi^{\alpha} \}$ of
the effective action of the $\phi^{\alpha}$-field
in Eq.\ref{eq:Skin3} at the fourth order, i.e. we may approximate
 \begin{eqnarray}
 S^{\prime}_{kin} \{ \phi^{\alpha} \} & \approx &
 S_{kin,3} \{ \phi^{\alpha} \} +
 S_{kin,4} \{ \phi^{\alpha} \}
 \nonumber
 \\
 & = &
 \frac{1}{3} \sum_{q_{1}  q_{2} q_{3} }
 \sum_{\alpha_{1} \alpha_{2} \alpha_{3} }
 U_{3} (
 q_1 \alpha_{1} q_{2} \alpha_{2} q_{3} \alpha_{3}  )
{\phi}^{\alpha_{1}}_{q_{1}}
 {\phi}^{\alpha_{2}}_{q_{2}}
 {\phi}^{\alpha_{3}}_{q_{3}}
 \nonumber
 \\
 & + &
 \frac{1}{4} \sum_{q_{1}  q_{2} q_{3} q_{4}}
 \sum_{\alpha_{1} \alpha_{2} \alpha_{3} \alpha_{4}}
 U_{4} (
 q_1 \alpha_{1} q_{2} \alpha_{2} q_{3} \alpha_{3}  q_{4} \alpha_{4} )
{\phi}^{\alpha_{1}}_{q_{1}}
 {\phi}^{\alpha_{2}}_{q_{2}}
 {\phi}^{\alpha_{3}}_{q_{3}}
 {\phi}^{\alpha_{4}}_{q_{4}}
 \label{eq:Skinprimeapprox}
 \; \; \; ,
 \end{eqnarray}
where the vertices $U_{3}$ and $U_{4}$ are defined in Eq.\ref{eq:Uvertex}.
According to the general formalism outlined above,
the bosonized kinetic energy
$S_{kin} \{ \tilde{\rho}^{\alpha} \}$
is obtained by calculating
the functional Fourier  transform of $S_{kin} \{ \phi^{\alpha } \}$.
Within one-loop approximation it is sufficient to retain
only the term $k=1$ in the linked cluster expansion \ref{eq:generalpert1}, so
that
 \begin{equation}
S_{kin} \left\{ \tilde{\rho}^{\alpha} \right\}
= S^{(0)}_{kin}
\left\{ \tilde{\rho}^{\alpha} \right\}
+
 \left<
 S_{kin}^{\prime \prime} \{ \phi^{\alpha} , \tilde{\rho}^{\alpha} \}
 \right>_{0}
 \; \; \; ,
 \label{eq:Skinrho1loop}
 \end{equation}
where
 \begin{eqnarray}
 S^{(0)}_{kin}
\left\{ \tilde{\rho}^{\alpha} \right\}
& =  &
 S_{kin,0}^{(0)} +
  \frac{1}{2} \sum_{q} \sum_{\alpha \alpha^{\prime}}
  \Gamma^{\alpha \alpha^{\prime}}_{q} \tilde{\rho}^{\alpha}_{-q}
\tilde{\rho}^{\alpha^{\prime}}_{q}
 \nonumber
 \\
& + &
 \frac{1}{3} \sum_{q_{1}  q_{2} q_{3} }
 \sum_{\alpha_{1} \alpha_{2} \alpha_{3} }
 \Gamma_{3}^{(0)} (
 q_1 \alpha_{1} q_{2} \alpha_{2} q_{3} \alpha_{3}  )
{\tilde{\rho}}^{\alpha_{1}}_{q_{1}}
 {\tilde{\rho}}^{\alpha_{2}}_{q_{2}}
 {\tilde{\rho}}^{\alpha_{3}}_{q_{3}}
 \nonumber
 \\
 & + &
 \frac{1}{4} \sum_{q_{1}  q_{2} q_{3} q_{4}}
 \sum_{\alpha_{1} \alpha_{2} \alpha_{3} \alpha_{4}}
 \Gamma_{4}^{(0)} (
 q_1 \alpha_{1} q_{2} \alpha_{2} q_{3} \alpha_{3}  q_{4} \alpha_{4} )
{\tilde{\rho}}^{\alpha_{1}}_{q_{1}}
 {\tilde{\rho}}^{\alpha_{2}}_{q_{2}}
 {\tilde{\rho}}^{\alpha_{3}}_{q_{3}}
 {\tilde{\rho}}^{\alpha_{4}}_{q_{4}}
 \label{eq:Skinapproxrho}
 \; \; \; ,
 \end{eqnarray}
with
 \begin{eqnarray}
 \Gamma_{3}^{(0)} (
 q_1 \alpha_{1} q_{2} \alpha_{2} q_{3} \alpha_{3}  )
 & = & - i \sum_{\alpha_{1}^{\prime} \alpha_{2}^{\prime} \alpha_{3}^{\prime} }
 U_{3} (
 q_1 \alpha_{1}^{\prime} q_{2} \alpha_{2}^{\prime} q_{3} \alpha_{3}^{\prime}  )
 \Gamma_{q_{1}}^{\alpha_{1}^{\prime} \alpha_{1}}
 \Gamma_{q_{2}}^{\alpha_{2}^{\prime} \alpha_{2}}
 \Gamma_{q_{3}}^{\alpha_{3}^{\prime} \alpha_{3}}
 \; \; \; ,
 \label{eq:Gamma03}
 \\
 \Gamma_{4}^{(0)} (
 q_1 \alpha_{1} q_{2} \alpha_{2} q_{3} \alpha_{3}  q_{4} \alpha_{4} )
 & = & \sum_{\alpha_{1}^{\prime} \alpha_{2}^{\prime} \alpha_{3}^{\prime}
\alpha_{4}^{\prime}}
 U_{4} (
 q_1 \alpha_{1}^{\prime} q_{2} \alpha_{2}^{\prime} q_{3} \alpha_{3}^{\prime}
q_{4} \alpha_{4}^{\prime})
 \Gamma_{q_{1}}^{\alpha_{1}^{\prime} \alpha_{1}}
 \Gamma_{q_{2}}^{\alpha_{2}^{\prime} \alpha_{2}}
 \Gamma_{q_{3}}^{\alpha_{3}^{\prime} \alpha_{3}}
 \Gamma_{q_{4}}^{\alpha_{4}^{\prime} \alpha_{4}}
 \label{eq:Gamma04}
 \; \; \; .
 \end{eqnarray}
The correction term due to one internal $\phi^{\alpha}$-loop is
 \begin{equation}
 \left<
 S_{kin}^{\prime \prime} \{ \phi^{\alpha} , \tilde{\rho}^{\alpha} \}
 \right>_{0} = S_{kin,0}^{(1)} + S_{kin,1}^{(1)} \left\{ \tilde{\rho}^{\alpha}
\right\}
 + S_{kin,2}^{(1)}
  \left\{ \tilde{\rho}^{\alpha} \right\}
 \label{eq:Skinloop}
 \; \; \; ,
 \end{equation}
where
 \begin{eqnarray}
S_{kin,0}^{(1)} & = &
\frac{3}{2} \sum_{q q^{\prime}}
 \sum_{\alpha_{1} \alpha_{2} \alpha_{3} \alpha_{4}}
 U_{4} ( - q \alpha_{1} , q \alpha_{2} , -q^{\prime} \alpha_{3} , q^{\prime}
\alpha_{4} )
 \Gamma_{q}^{\alpha_{2} \alpha_{1}}
 \Gamma_{q^{\prime}}^{\alpha_{4} \alpha_{3}}
 \; \; \; ,
 \label{eq:Skin01}
 \\
 S_{kin,1}^{(1)}
  \left\{ \tilde{\rho}^{\alpha} \right\}
 &  = & \sum_{\alpha} \Gamma_{1}^{(1)} ( \alpha ) \tilde{\rho}^{\alpha}_{0}
 \; \; \; ,
 \label{eq:Skin11}
 \\
 S_{kin,2}^{(1)}
  \left\{ \tilde{\rho}^{\alpha} \right\}
 &  = &
  \frac{1}{2} \sum_{q} \sum_{\alpha \alpha^{\prime}}
  \Gamma^{(1)}_{2} ( - q {\alpha} , q \alpha^{\prime} )
\tilde{\rho}^{\alpha}_{-q}
  \tilde{\rho}^{\alpha^{\prime}}_{q}
 \label{eq:Skin12}
 \; \; \; ,
 \end{eqnarray}
with
 \begin{eqnarray}
 \Gamma_{1}^{(1)} ( \alpha ) & = & i \sum_{q}
 \sum_{\alpha_{1} \alpha_{2} \alpha_{3} }
 U_{3} ( -q \alpha_{1} , q \alpha_{2} , 0 \alpha_{3} ) \Gamma_{q}^{ \alpha_{2}
\alpha_{1} }
 \Gamma_{0}^{\alpha \alpha_{3} }
 \; \; \; ,
 \label{eq:Gamma11def}
 \\
  \Gamma^{(1)}_{2} ( - q {\alpha} , q \alpha^{\prime} )
  & = &
  - 3 \sum_{q^{\prime}}
 \sum_{\alpha_{1} \alpha_{2} \alpha_{3} \alpha_{4}}
 U_{4} ( - q \alpha_{1} , q \alpha_{2} , -q^{\prime} \alpha_{3} , q^{\prime}
\alpha_{4} )
 \Gamma_{q}^{\alpha \alpha_{1}}
 \Gamma_{q}^{\alpha_{2} \alpha^{\prime}}
 \Gamma_{q^{\prime}}^{\alpha_{4} \alpha_{3}}
 \label{eq:Gamma21def}
 \; \; \; .
 \end{eqnarray}
Recall that the superscript $(1)$ indicates that these terms contain one
internal bosonic loop.
Thus, within the one-loop approximation the constant in Eq.\ref{eq:Skinn}
is $S_{kin,0} =  S_{kin,0}^{(0)} + S_{kin,0}^{(1)}$,
and the vertices $\Gamma_{n}$ in Eq.\ref{eq:Gammank} are approximated by
 \begin{eqnarray}
 \Gamma_{1} ( q \alpha ) & = & \Gamma_{1}^{(1)} ( q \alpha )
 \; \; \; ,
 \\
 \Gamma_{2} ( -q \alpha , q \alpha^{\prime} ) & = & \Gamma_{q}^{\alpha
\alpha^{\prime} }
 + \Gamma_{2}^{(1)} ( -q \alpha , q \alpha^{\prime} )
 \; \; \; ,
 \\
 \Gamma_{3} ( q_{1} \alpha_{1} q_{2} \alpha_{2} q_{3} \alpha_{3} )
 & = &
 \Gamma_{3}^{(0)} ( q_{1} \alpha_{1} q_{2} \alpha_{2} q_{3} \alpha_{3} )
 \; \; \; ,
 \\
 \Gamma_{4} ( q_{1} \alpha_{1} q_{2} \alpha_{2} q_{3} \alpha_{3} q_{4}
\alpha_{4} )
 & = &
 \Gamma_{4}^{(0)} ( q_{1} \alpha_{1} q_{2} \alpha_{2} q_{3} \alpha_{3} q_{4}
\alpha_{4} )
 \; \; \; ,
 \end{eqnarray}
and all $\Gamma_{n}$ with $n \geq 5$ are set equal to zero.
The term with $\Gamma_{1}$ can be ignored for a calculation of correlation
functions at finite $q$, because
it involves only the $q=0$ component of the density fields. Furthermore,
at the level of a one-loop calculation we may also ignore
the term $\Gamma_{3}^{(0)}$, because the Gaussian expectation value of a
product of three
$\tilde{\rho}^{\alpha}$-fields vanishes.
Combining the relevant contributions from the kinetic energy with the
interaction contribution, we finally arrive at the following
effective action
 \begin{eqnarray}
 {S}_{eff} \left\{ \tilde{\rho}^{\alpha} \right\}   & = &
 \frac{1}{2} \sum_{q} \sum_{\alpha \alpha^{\prime}}
 \left[  [ \underline{\tilde{f}}_{q} ]^{ \alpha \alpha^{\prime} }
  + \Gamma_{q}^{\alpha \alpha^{\prime}}
  \right]
 \tilde{\rho}^{\alpha}_{-q}
 \tilde{\rho}^{\alpha^{\prime}}_{q}
 +
  \frac{1}{2} \sum_{q} \sum_{\alpha \alpha^{\prime}}
  \Gamma^{(1)}_{2} ( - q {\alpha} , q \alpha^{\prime} )
\tilde{\rho}^{\alpha}_{-q}
 \tilde{\rho}^{\alpha^{\prime}}_{q}
  \nonumber
  \\
  & + &
 \frac{1}{4} \sum_{q_{1}  q_{2} q_{3} q_{4}}
 \sum_{\alpha_{1} \alpha_{2} \alpha_{3} \alpha_{4}}
 \Gamma_{4}^{(0)} (
 q_1 \alpha_{1} q_{2} \alpha_{2} q_{3} \alpha_{3}  q_{4} \alpha_{4} )
{\tilde{\rho}}^{\alpha_{1}}_{q_{1}}
 {\tilde{\rho}}^{\alpha_{2}}_{q_{2}}
 {\tilde{\rho}}^{\alpha_{3}}_{q_{3}}
 {\tilde{\rho}}^{\alpha_{4}}_{q_{4}}
 \label{eq:Seff1loop}
 \; \; \; .
 \end{eqnarray}
We emphasize that this effective action is only good for the purpose of
calculating the
one-loop corrections to the Gaussian approximation.
At two-loop order one should also retain the terms with $\Gamma_{3}$ and
$\Gamma_{6}$.
The last two terms in Eq.\ref{eq:Seff1loop} contain the one-loop corrections to
the
non-interacting boson approximation for the bosonized collective density
fluctuations.
In the limit of long wavelengths we may again write down an equivalent
effective Hamiltonian
of canonically quantized bosons by using the substitution \ref{eq:rhobscale}.
However, we shall not even bother writing down this complicated expression,
because
this mapping is only valid at long wavelengths and high densities, and does not
lead to any simplification.
For all practical purposes the parametrization
in terms of the collective density-field $\tilde{\rho}^{\alpha}$ is
superior.
We shall now use this parametrization to calculate the
leading correction to the bosonic propagator and determine in this way
the hidden small parameter
which controls the range of validity of the Gaussian approximation.

\subsection{First order self-energy and hidden small parameter}
\label{subsec:hidden}

Let us define a dimensionless proper (or irreducible) self-energy matrix
$\underline{\Sigma}^{\ast}_{q}$
via
 \begin{equation}
 \left< \tilde{\rho}^{\alpha}_{q} \tilde{\rho}^{\alpha^{\prime} }_{q^{\prime}}
\right>
 = \delta_{q + q^{\prime} , 0}
 \left[ \left[   \underline{\tilde{f}}_{q} + \underline{\Gamma}_{q} -
 \underline{\Sigma}^{\ast}_{q} \right]^{-1}
 \right]^{\alpha \alpha^{\prime}}
 \label{eq:rhoprop}
 \; \; \; ,
 \end{equation}
where the probability distribution for the average is determined by
the exact effective action $S_{eff} \{ \tilde{\rho}^{\alpha} \}$, see
Eqs.\ref{eq:Pirho} and \ref{eq:Seffrhodef}.
{}From Eq.\ref{eq:rhopropgauss} it is clear
that the self-energy $\underline{\Sigma}^{\ast}_{q}$ contains by definition all
corrections to the
RPA.
Introducing the exact proper polarization matrix $\underline{\Pi}^{\ast} (q)$
via
 \begin{equation}
 [ \underline{\Pi}^{\ast } ( q )    ]^{-1}
  =  [ \underline{\Pi}_{0} (q ) ]^{-1} - \frac{V}{\beta}
\underline{\Sigma}^{\ast}_{q}
  \; \; \; ,
 \label{eq:Piast}
  \end{equation}
the exact total density-density correlation function can be written as
 \begin{equation}
 \Pi ( q ) = \sum_{\alpha \alpha^{\prime}}
 \left[ \left[  [ \underline{\Pi}^{\ast  } ( q) ]^{-1} +  \underline{f}_{q}
 \right]^{-1} \right]^{\alpha \alpha^{\prime} }
 \; \; \; .
 \label{eq:Piexact}
 \end{equation}
If all matrix elements of $\underline{f}_{q}$ are identical and equal
to $f_{\bf{q}}$, we may repeat the manipulations in
Eqs.\ref{eq:Ptotdecompose1}-\ref{eq:Pi0tot}, so that
Eq.\ref{eq:Piexact} reduces to\cite{Mahan81}
 \begin{equation}
 \Pi ( q ) = \frac{\Pi^{\ast} ( q )}{1 + f_{\bf{q}} \Pi^{\ast} ( q ) }
 \; \; \; ,
 \label{eq:Piasttot}
 \end{equation}
where
 \begin{equation}
 \Pi^{\ast} ( q ) = \sum_{\alpha \alpha^{\prime}}
 [ \underline{\Pi}^{\ast} ( q ) ]^{\alpha \alpha^{\prime}}
 \; \; \; ,
 \label{eq:Piglobprop}
 \end{equation}
is the total proper polarization.
With the help of the dielectric function
 \begin{equation}
 \epsilon ( q ) = 1 + f_{\bf{q}} \Pi^{\ast} ( q )
 \; \; \; ,
 \label{eq:dielectricdef}
 \end{equation}
Eq.\ref{eq:Piasttot} can also be written as
 \begin{equation}
 \Pi ( q ) = \frac{\Pi^{\ast} ( q )}{ \epsilon( q ) }
 \; \; \; .
 \label{eq:Piasttot2}
 \end{equation}

\vspace{7mm}

To first order in an expansion in the number of bosonic loops,
we simply have to add the two diagrams
shown in Fig.\ref{fig:self1}.
Because we have symmetrized the vertices, the
loop diagram in Fig.\ref{fig:self1} has a combinatorial factor of
three, so that to first order $\underline{\Sigma}^{\ast}_{q} \approx
\underline{\Sigma}^{(1)}_{q}$, with
 \begin{equation}
 [ \underline{\Sigma}_{q}^{(1)} ]^{\alpha \alpha^{\prime} }
 =  - \Gamma_{2}^{(1)} ( -q \alpha , q \alpha^{\prime} )
 -   3 \sum_{q^{\prime}} \sum_{\alpha_{3} \alpha_{4} }
 \Gamma_{4}^{(0)} ( - q \alpha , q
 \alpha^{\prime} , -q^{\prime} \alpha_{3} ,q^{\prime} \alpha_{4} )
 \left[ \left[  \underline{\tilde{f}}_{q^{\prime}} +
 \underline{\Gamma}_{q^{\prime}} \right]^{-1} \right]^{\alpha_{4} \alpha_{3}}
 \label{eq:self1}
 \; \; \; .
 \end{equation}
Using the definitions of $\Gamma_{2}^{(1)}$ and $\Gamma_{4}^{(0)}$
(see Eqs.\ref{eq:Gamma21def} and \ref{eq:Gamma04}),
it is easy to show that Eq.\ref{eq:self1} can also be written as
 \begin{eqnarray}
 [ \underline{\Sigma}_{q}^{(1)} ]^{\alpha \alpha^{\prime} }
 & = &  3
 \sum_{q^{\prime}} \sum_{\alpha_{1} \alpha_{2} \alpha_{3} \alpha_{4} }
 U_{4} ( - q \alpha_{1} , q
 \alpha_{2} , -q^{\prime} \alpha_{3} , q^{\prime} \alpha_{4} )
 \Gamma_{q}^{\alpha \alpha_{1}}
 \Gamma_{q}^{\alpha_{2} \alpha^{\prime}}
 \nonumber
 \\
 & \times &
 \left[
 \underline{\Gamma}_{q^{\prime}} -
 \underline{\Gamma}_{q^{\prime}}
 \left[  \underline{\tilde{f}}_{q^{\prime}} + \underline{\Gamma}_{q^{\prime}}
\right]^{-1}
 \underline{\Gamma}_{q^{\prime}}
 \right]^{ \alpha_{4} \alpha_{3} }
 \label{eq:self2}
 \; \; \; .
 \end{eqnarray}
A simple manipulation of the matrix in the last line  gives
 \begin{equation}
 \underline{\Gamma}_{q} -
 \underline{\Gamma}_{q}
 \left[  \underline{\tilde{f}}_{q} + \underline{\Gamma}_{q} \right]^{-1}
 \underline{\Gamma}_{q}
  =
   \underline{\tilde{f}}_{q} \left[ 1 + \underline{\Gamma}_{q}^{-1}
  \underline{\tilde{f}}_{q} \right]^{-1}
 =
 \frac{ \beta}{V}  \underline{f}^{RPA}_{q}
 \label{eq:frpamani}
 \; \; \; ,
 \end{equation}
where the matrix $\underline{f}^{RPA}_{q}$ is defined by
 \begin{equation}
 \underline{f}^{RPA}_{q} =
 \underline{{f}}_{q}
 \left[ 1 + \underline{\Pi}_{0} (q)  \underline{f}_{q}
 \right]^{-1}
\label{eq:frpadef}
\; \; \; .
\end{equation}
It follows that Eq.\ref{eq:self2} reduces to
 \begin{equation}
 [ \underline{\Sigma}_{q}^{(1)} ]^{\alpha \alpha^{\prime} }
  =   3 \frac{ \beta}{V}
 \sum_{q^{\prime}} \sum_{\alpha_{1} \alpha_{2} \alpha_{3} \alpha_{4} }
 \Gamma_{q}^{\alpha \alpha_{1}}
 \Gamma_{q}^{\alpha_{2} \alpha^{\prime}}
 U_{4} ( - q \alpha_{1} , q
 \alpha_{2} , -q^{\prime} \alpha_{3} , q^{\prime} \alpha_{4} )
 \left[ \underline{f}^{RPA}_{q^{\prime} }
 \right]^{ \alpha_{4} \alpha_{3} }
 \label{eq:self3}
 \; \; \; .
 \end{equation}
Note that $\underline{\Sigma}_{q}^{(1)}$ is proportional to the
RPA-screened interaction and vanishes in the non-interacting limit,
as it should. Eq.\ref{eq:self3} is the general result for the
leading correction to the Gaussian approximation in an arbitrary
patch-geometry,
and for an arbitrary interaction matrix $\underline{f}_{q}$.

To make further progress,
we shall now restrict ourselves to the high-density limit
and assume that the interaction is negligibly small for wave-vectors
larger than a cutoff
$ q_{c} \ll min \{ \Lambda , \lambda \} \ll k_{F}$.
Choosing also the magnitude of the external wave-vector ${\bf{q}}$
in Eq.\ref{eq:self3} small
compared with the characteristic size of the patches,
the {\it{diagonal-patch approximation}} $(A1)$ is justified, so that
$\underline{\Gamma}_{q}$
and
 $U_{4} ( - q \alpha_{1} , q
 \alpha_{2} , -q^{\prime} \alpha_{3} , q^{\prime} \alpha_{4} ) $
are diagonal in the patch-indices, see Eqs.\ref{eq:Gammalong}
and \ref{eq:Undiag}.
Then Eq.\ref{eq:self3} reduces to
 \begin{equation}
 [ \underline{\Sigma}_{q}^{(1)} ]^{\alpha \alpha^{\prime} }
 = \delta^{\alpha \alpha^{\prime}} \frac{ \beta}{V \nu^{\alpha}}
 \left( \frac{  {\bf{v}}^{\alpha} \cdot {\bf{q}} - i \omega_{m} }
 {  {\bf{v}}^{\alpha} \cdot {\bf{q}} } \right)^2 A^{\alpha}_{q}
 \; \; \; ,
 \label{eq:Sigma1diag}
 \end{equation}
where the dimensionless function $A^{\alpha}_{q}$ is given by
 \begin{equation}
 A^{\alpha}_{q} =  \frac{ 3 \beta}{V \nu^{\alpha} }
 \sum_{q^{\prime}} U^{\alpha}_{4} ( - q  , q , - q^{\prime} , q^{\prime}  )
 \left[ \underline{f}^{RPA}_{q^{\prime}} \right]^{\alpha \alpha}
 \label{eq:Aalphadef}
 \; \; \; .
 \end{equation}
We conclude that in the high-density- and long wavelength limit
we have to first order in the screened interaction
 \begin{equation}
  \Gamma_{q}^{\alpha \alpha^{\prime} }
 - [ \underline{ \Sigma }_{q}^{(1)} ]^{\alpha \alpha^{\prime} }
 =
 \frac{\beta}{V}
 [ [ \underline{\Pi}^{\ast } ( q ) ]^{-1} ]^{\alpha \alpha^{\prime}}
 = \frac{\beta}{V} \frac{\delta^{\alpha \alpha^{\prime}} }{\Pi^{\ast  \alpha} (
q ) }
 \; \; \; ,
 \label{eq:Piast1}
 \end{equation}
with
 \begin{equation}
 \frac{1}{\Pi^{\ast  \alpha} ( q ) }
 =  \frac{
  ( 1 - A^{\alpha}_{q} )
   {\bf{v}}^{\alpha} \cdot {\bf{q}}
  - ( 1 - 2 A^{\alpha}_{q} ) i \omega_{m} - A^{\alpha}_{q}
  \frac{ (i \omega_{m} )^2}{  {\bf{v}}^{\alpha} \cdot {\bf{q}} }
  }
  { \nu^{\alpha} \;
   {\bf{v}}^{\alpha} \cdot {\bf{q}} }
  \label{eq:Pi1def}
  \; \; \; .
  \end{equation}
Comparing this expression with Eq.\ref{eq:Gammalong}, we conclude that
the non-interacting boson approximation is quantitatively correct provided the
condition
 \begin{equation}
 | A^{\alpha}_{q} | \ll 1
 \end{equation}
is satisfied for all patches $\alpha$,
because then the corrections to the propagator of the locally defined
collective
density field $\tilde{\rho}^{\alpha}_{q}$ are small.

For simplicity, let us now assume that all matrix elements
of $\underline{f}_{q}$ are identical.
Using the same procedure as in Eq.\ref{eq:Neumann}, it is then easy to show
that
$[\underline{f}^{RPA}_{q} ]^{\alpha \alpha} = f^{RPA}_{q}$ is also independent
of the
patch label.
{}From the general definition of the vertices $U_{n}$ in Eq.\ref{eq:Uvertex}
we obtain in this case
 \begin{eqnarray}
 A^{\alpha}_{q}
 & = &
 \frac{1}{\nu^{\alpha}  \beta V }
 \sum_{k}
 \Theta^{\alpha} ( {\bf{k}} )
 \left\{     G_{0} ( k) \Sigma_{F}^{(1)} ( k )  G_{0} ( k ) [ G_{0} ( k + q )
+ G_{0} ( k- q ) ]
 \right.
 \nonumber
 \\
 & & +
 \left. \frac{1}{2}
  G_{0} ( k )   [ \Gamma_{F}^{(1) } ( k , q )  G_{0} ( k + q )
  + \Gamma_{F}^{(1)} ( k , -q ) G_{0}  ( k -q ) ]
 \right\}
 \label{eq:Aalphaqres}
 \; \; \; ,
 \end{eqnarray}
with
 \begin{eqnarray}
 \Sigma_{F}^{(1)} ( k ) & = &
 \frac{1}{\beta V } \sum_{q^{\prime}}
 f^{RPA}_{{q}^{\prime}} G_{0} ( k+q^{\prime} )
 \; \; \; ,
 \label{eq:sigmaF1}
 \\
 \Gamma_{F}^{(1)} ( k , q) & = &
 \frac{1}{  \beta V } \sum_{q^{\prime}}
 f^{RPA}_{{q}^{\prime}}
 G_{0} ( k+q^{\prime})  G_{0} ( k + q^{\prime} + q )
 \label{eq:gammaF1}
 \; \; \; .
 \end{eqnarray}
Note  that $A^{\alpha}_{-q} = A^{\alpha}_{q}$ due to the symmetrization of the
vertex $U_{4}$ with respect to the interchange of any two labels.
It is now obvious that the vertices of our effective bosonic
action automatically contain the relevant self-energy and vertex corrections
of the underlying fermionic problem\cite{Geldart70}.
The first term in Eq.\ref{eq:Aalphaqres} corresponds to the
two self-energy corrections to the non-interacting polarization
bubble shown in Fig.\ref{fig:bubblecor} (a) and (b), while the last term
is due to the vertex correction shown in Fig.\ref{fig:bubblecor} (c).

\vspace{7mm}

In order to determine the range of validity of the non-interacting boson
approximation,
we have to
calculate the dependence of  $A^{\alpha}_{q}$ on the various parameters in the
problem.
In the limit of long wavelengths and low energies,
it is to leading order in $|  {\bf{v}}^{\alpha} \cdot {\bf{q}}|$ and
$| \omega_{m} |$  consistent to replace in  Eq.\ref{eq:Pi1def}
$A^{\alpha}_{q} \rightarrow A^{\alpha}_{0}$.
Actually, the $q \rightarrow 0$ limit of $A^{\alpha}_{q}$ should be taken in
such a way that
the ratio $ \frac{ i \omega_{m}  }{   {\bf{v}}^{\alpha} \cdot {\bf{q}}  }$ is
held constant, because
in this case we obtain the low-energy behavior of
$A^{\alpha}_{q}$ close to the poles of the Gaussian propagator.
However, since we are only interested in the order of magnitude of
$A^{\alpha}_{q}$ for small $| \omega_{m} |$ and
$|{\bf{q}} | $, it is sufficient to consider the "${\bf{q}}$-limit"
$A^{\alpha}_{0} = \lim_{\bf{q} \rightarrow 0} [ \lim_{\omega_{m} \rightarrow 0
} A^{\alpha}_{q} ]$.
For simplicity, let us
now ignore the frequency-dependence of the RPA-interaction.
This amounts to the static approximation for the dielectric constant, which
seems reasonable to obtain the correct order of magnitude of $A^{\alpha}_{0}$.
The "${\bf{q}}$-limit" is obtained by
setting $q=0$ under the summation sign and performing the Matsubara sums before
doing the wave-vector integrations.
For $\beta \rightarrow \infty$ we obtain
 \begin{equation}
 A^{\alpha}_{0} =  \frac{1}{\nu^{\alpha} V^2}
 \sum_{ {\bf{q}}^{\prime} {\bf{k}} }
 \Theta^{\alpha} ( {\bf{k}} )
 f^{RPA}_{\bf{q}^{\prime}}
 \left\{ f (  \xi_{\bf{k+q^{\prime}}} )
 \frac{ \partial^{2}}{\partial \mu^2} f ( \xi_{\bf{k}}  )
 + \frac{\partial}{\partial \mu } f ( \xi_{ \bf{k} + \bf{q}^{\prime} }  )
 \frac{\partial}{\partial \mu } f ( \xi_{ \bf{k} }  )
 \right\}
 \label{eq:A0res1}
 \; \; \; .
 \end{equation}
Because the
${\bf{k}}$-sum extends over the entire box $K^{\alpha}$ and
by assumption the ${\bf{q}^{\prime}}$-sum is cut off by the interaction
at $q_{c} \ll min \{ \Lambda , \lambda \}$,
we may set $\xi_{\bf{k} + \bf{q}^{\prime} } \approx \xi_{\bf{k}}$ in the
Fermi-functions
of Eq.\ref{eq:A0res1}. Then the summations factorize,
and we obtain
 \begin{equation}
 A^{\alpha}_{0} =
 \left[ \frac{1}{V} \sum_{\bf{q} }
 f^{RPA}_{\bf{q}} \right]
  \frac{1}{ \nu^{\alpha} } \int_{- \infty}^{\infty} d \xi
 \nu^{\alpha} ( \xi )
  \left[ f ( \xi ) f^{\prime \prime} ( \xi )
  + f^{\prime} ( \xi ) f^{\prime} ( \xi )  \right]
  \label{eq:A0res2}
  \; \; \; ,
  \end{equation}
where $\nu^{\alpha} ( \xi )$ is the
energy dependent patch density of states,
 \begin{equation}
 \nu^{\alpha} ( \xi ) = \frac{1}{V} \sum_{\bf{k}}
 \Theta^{\alpha} ( {\bf{k}} ) \delta (  \xi - \xi_{\bf{k}}  )
 \label{eq:nualphae}
 \; \; \; .
 \end{equation}
Note that from the definition of $\nu^{\alpha}$ in Eq.\ref{eq:nualphadef}
 \begin{equation}
 \nu^{\alpha} = \int_{- \infty}^{\infty} d \xi
 \nu^{\alpha} ( \xi ) \left[ -  f^{\prime} ( \xi )
 \right]
 \label{eq:nunu}
 \; \; \; .
 \end{equation}
Integrating in Eq.\ref{eq:A0res2} by parts, we obtain in the limit $\beta
\rightarrow \infty$
 \begin{eqnarray}
  \lefteqn{  \int_{- \infty}^{\infty} d \xi
 \nu^{\alpha} ( \xi )
  \left[ f ( \xi ) f^{\prime \prime} ( \xi )
  +  f^{\prime} ( \xi ) f^{\prime} ( \xi)  \right] = }
  \\
  &   &
  \int_{- \infty}^{\infty} d \xi
 \nu^{\alpha} ( \xi ) \frac{\partial}{\partial \xi}
  \left[ f ( \xi ) f^{\prime} ( \xi ) \right]
 = \frac{1}{ 2 } \frac{ \partial \nu^{\alpha}}{ \partial \mu }
 \label{eq:intebyparts1}
 \; \; \; .
 \end{eqnarray}
Because by assumption $f^{RPA}_{\bf{q}}$
becomes negligibly small for $|{\bf{q}}  | \geq q_{c}$,
the first factor in Eq.\ref{eq:A0res2}
is for $V \rightarrow \infty$
 \begin{equation}
  \frac{1}{V} \sum_{\bf{q} }
 f^{RPA}_{\bf{q}} \propto {f}^{RPA}_{0} q_{c}^d
 \label{eq:firstfac}
 \; \; \; .
 \end{equation}
Ignoring a numerical factor of the order of unity,
the final result for $A^{\alpha}_{0}$ can be written as
 \begin{equation}
A^{\alpha}_{0} \approx \frac{ q_{c}^d {f}^{RPA}_{0} }{\mu } C^{\alpha}
 \label{eq:Afinal1}
 \; \; \; ,
 \end{equation}
where the dimensionless parameter
 \begin{equation}
 C^{\alpha} = \frac{ \mu}{\nu^{\alpha}} \frac{ \partial \nu^{\alpha} }{\partial
\mu }
 = \frac{ \mu \frac{ \partial^{2} N^{\alpha}_{0}}{\partial \mu^2} }
 { \frac{ \partial N^{\alpha}_{0} }{\partial \mu }}
 \label{eq:g1def}
 \end{equation}
is a measure for the {\it{local curvature of the Fermi surface in patch
$\tilde{K}^{\alpha}_{\Lambda}$}}.
For sufficiently smooth Fermi surfaces the patch density of states
$\nu^{\alpha}$ is
proportional to
$ \Lambda^{d-1} $.
However,
the cutoff dependence cancels in Eq.\ref{eq:g1def}, because it
appears in the enumerator as well as in the denominator.
Therefore $C^{\alpha}$ is a cutoff-independent quantity.
If we linearize the
energy dispersion in patch $\tilde{K}^{\alpha}_{\lambda}$, then the local
density of states
is replaced by a constant independent of $\mu$, so that
$C^{\alpha}$ vanishes in this case.
Then there is no correction to the Gaussian approximation.
Of course, we already know from the closed loop theorem that
the Gaussian approximation is exact if in addition
to the diagona-patch approximation $(A1)$
the energy dispersion is linearized for all patches.
As usual, we introduce the
dimensionless interaction $F^{RPA}_{0} = \nu f^{RPA}_{0}$,
which measures the strength of the potential energy relative to the
kinetic energy.
Note that for regular interactions
$F^{RPA}_{0} \approx \frac{ \nu f_{0}}{1 + \nu f_{0}}$.
Using the fact to the global density of states is in $d$
dimensions proportional to $ k_{F}^{d-2}$,
we conclude that in the high-density limit the Gaussian approximation is
quantitatively correct provided the condition
 \begin{equation}
 | A^{\alpha}_{0} | \equiv  \left( \frac{ q_{c}}{k_{F} } \right)^{d} |
{F}^{RPA}_{0} |  | C^{\alpha}| \ll 1
 \label{eq:gausscorrect}
 \end{equation}
is satisfied. This is the main result of this section.

The appearance of three parameters that control the accuracy
of the Gaussian approximation has a very simple intuitive interpretation.
First of all, if at all points on the Fermi surface the
curvature is intrinsically small
(i.e. $| C^{\alpha} | \ll 1$ for all patches $\alpha$)
then the corrections to the linearization of the energy dispersion are
small, and hence the Gaussian approximation becomes accurate.
Note that in the one-dimensional Tomonaga-Luttinger model
$C^{\alpha} = 0$,
because the energy dispersion is linear by definition.
However, for realistic energy dispersions of the form
$\epsilon_{\bf{k}} = \frac{ {\bf{k}}^{2}}{2 m}$ the
dimensionless curvature parameter
$C^{\alpha}$ is of the order of unity.
But even then the
Gaussian approximation is accurate, provided the nature of the interaction
is such that it involves only small momentum transfers.
This is also intuitively obvious, because in this case
the scattering processes probe only a thin shell around the
Fermi surface and do not feel the deviations from linearity.
Finally, it is clear that also the strength of the effective interaction should
determine the range of validity of Gaussian approximation,
because in the limit that
the strength of the interaction approaches zero
all corrections to the Gaussian approximation vanish.

\vspace{7mm}

Finally, we would like to discuss a subtlety associated with
the choice of the radial patch cutoff $\lambda$ and the picture of having
integrated
out all degrees of freedom outside a thin shell around the
Fermi surface.
Consider the three dimensional Coulomb interaction
in the homogeneous electron gas, which in the static screening approximation
corresponds in the regime
$|{\bf{q}} | \ll k_{F}$ to
 \begin{equation}
 F_{\bf{q}}^{RPA} = \frac{ \kappa^2 }{ {\bf{q}}^2 + \kappa^2 }
 \label{eq:fqcb}
 \; \; \; ,
 \end{equation}
where $\kappa = [ 4 \pi e^2 \nu ]^{1/2}$ is the Thomas-Fermi screening
wave-vector.
For $|{\bf{q}} | \approx k_{F}$ the Coulomb potential
reduces to a constant of the order of $\frac{e^{2}}{k_{F}^2}$.
If we introduce
a radial patch cutoff $\lambda \approx \kappa$ and {\it{define}} our bare model
such that
it contains only the degrees of freedom close to the
Fermi surface (see Sec.\ref{sec:intro}), then
we see from Eq.\ref{eq:fqcb} that we must choose $q_{c} \approx \kappa$ and
$F^{RPA}_{0} \approx 1$, so that Eq.\ref{eq:gausscorrect} gives
 \begin{equation}
  A^{\alpha}_{0}  \approx \left( \frac{\kappa}{k_{F}} \right)^{3}
 \label{eq:Acb1}
 \;\; \; , \; \; \; \lambda \approx \kappa
 \; \; \; ,
 \end{equation}
where we have ignored numerical
constants of the order of unity.
Note that $C^{\alpha} = O(1)$ for free electrons in $d=3$.
On the other hand, if we do not introduce
a radial patch cutoff but  treat also the high-energy degrees of freedom
explicitly, then
our simple estimate
in Eq.\ref{eq:gausscorrect} is not valid,
because $F^{RPA}_{\bf{q}}$ in Eq.\ref{eq:fqcb} does not fall off sufficiently
fast, so that
the summations in Eq.\ref{eq:A0res1} do not factorize.
Substituting Eq.\ref{eq:fqcb}
into \ref{eq:A0res1}, it is not difficult to show that
in this case
the value of $A^{\alpha}_{0}$
is essentially determined by the short-wavelength regime
$\kappa \leq | {\bf{q}} | \leq k_{F}$, and that it is given by
 \begin{equation}
 A^{\alpha}_{0} \approx \left( \frac{ \kappa }{k_{F}} \right)^2
 \;\; \; , \; \; \; \lambda = \infty
 \; \; \; .
 \label{eq:Acb2}
 \end{equation}
Although in both cases the corrections to the Gaussian approximation are small
at high densities (where $\kappa \ll k_{F}$), the corrections are
smaller by a factor of $\kappa / k_{F} $ in the model with a radial patch
cutoff.
This is evidently due to the fact that in the
cutoff model the high-energy degrees of freedom have already
been taken into account implicitly via the re-definition of the
parameters in the Gaussian approximation, so that only the much smaller
number of degrees of freedom in a thin shell around the Fermi surface can give
rise
to corrections.
However, our approach does not depend on the
linearization of the energy dispersion, so that the radial cutoff can be
removed
and also the high-energy degrees of freedom can be explicitly
taken into account. This enables us to use our bosonization
approach as a  basis
to develop a new systematic method of calculating corrections to the RPA.
We shall discuss this method in the following subsection.

\subsection{Calculating corrections to the RPA via bosonization}
\label{subsec:diagram}

The corrections to the RPA are usually expressed in terms of the
dynamic local field factor $g ( q ) $,
which is defined by writing the exact irreducible
polarization (see Eq.\ref{eq:Piglobprop}) in the
form\cite{Mahan81}-\cite{Gorabchenko89}
 \begin{equation}
 \Pi^{\ast} ( q ) = \frac{ \Pi_{0} (q)}{ 1 - g ( q ) \Pi_{0} ( q ) }
 \label{eq:gdyndef}
 \; \; \; .
 \end{equation}
Note that this equation can also be written as
 \begin{equation}
 [ {\Pi^{\ast} (q) }]^{-1} = [{\Pi_{0} ( q ) }]^{-1} - g ( q )
 \label{eq:Dysonpiglob}
 \; \; \; ,
 \end{equation}
which has the structure $G^{-1} = G_{0}^{-1} - \Sigma$, i.e. it resembles the
Dyson equation for the single-particle Greens-function of a bosonic problem,
with the
proper polarization and the local field factor playing the role of the exact
Greens-function
and the irreducible self-energy.
This analogy is well known\cite{Mahan81}-\cite{Gorabchenko89}, although it
seems that it has not been
thoroughly exploited as a guide to develop systematic methods to calculate
corrections
to the RPA. Such a method naturally emerges from our bosonization approach.
Defining a local-field correction matrix
 \begin{equation}
 \underline{g} ( q ) = \frac{V}{\beta} \underline{\Sigma}^{\ast}_{q}
 \; \; \; ,
 \label{glocalmatrix}
 \end{equation}
our matrix Dyson equation \ref{eq:Piast} for the patch proper polarization
takes the form
 \begin{equation}
 [ \underline{\Pi}^{\ast } ( q )    ]^{-1}
  =  [ \underline{\Pi}_{0} (q ) ]^{-1} -  \underline{g} ( q )
  \; \; \; .
 \label{eq:Piast2}
  \end{equation}
Comparison with Eq.\ref{eq:Dysonpiglob} shows that
the matrix elements $[ \underline{\Sigma}^{\ast}_{q} ]^{\alpha
\alpha^{\prime}}$
of the irreducible self-energy of our
effective bosonic model can be identified physically with
generalized local field corrections $ [\underline{g}(q) ]^{\alpha
\alpha^{\prime}}$,
which differentiate between the contributions from the
various patches.
For simplicity let us assume that the diagonal-patch approximation $(A1)$
is justified, so that Eq.\ref{eq:Piast2} reduces to an
equation for the diagonal elements, which can be written as
 \begin{eqnarray}
 \Pi^{\ast \alpha} ( q ) & = & \frac{ \Pi_{0}^{\alpha} ( q ) }{ 1 -
 g^{\alpha} ( q ) \Pi_{0}^{\alpha} ( q) }
 \nonumber
 \\
 & \approx &
 \Pi_{0}^{\alpha} ( q ) +
 \Pi_{0}^{\alpha} ( q )
 g^{ \alpha} (q) \Pi_{0}^{\alpha} ( q)
 + \ldots
 \; \; \; .
 \label{eq:Dysonexp}
 \end{eqnarray}
Here $ \Pi^{\ast \alpha} ( q ) = [ \underline{\Pi}^{\ast} ( q ) ]^{\alpha
\alpha }$,
$ g^{\alpha} (q ) = [ \underline{g} (q) ]^{\alpha \alpha }$, and
$\Pi_{0}^{\alpha} ( q )$ is at long wavelengths given
in Eq.\ref{eq:Pilong}.
The difference between our approach, which is based on the perturbative
calculation of the {\it{inverse}} proper polarization, and the
naive perturbative approach is now clear.
In the latter method the corrections to the proper polarization bubble are
determined by
direct expansion of $\Pi^{\ast} ( q )$ in powers of the
interaction\cite{Holas79}.
Such a procedure does not
correspond to the perturbative calculation of the {\it{self-energy}}, but
is equivalent to a direct expansion
of the {\it{Greens-function}}.
To first order, only the leading correction in the
expansion of the Dyson equation
(the second line in Eq.\ref{eq:Dysonexp}) is kept in this method,
so that the total proper polarization is approximated by
 \begin{eqnarray}
 \Pi^{\ast} ( q ) & \approx & \sum_{\alpha}
 \left[
 \frac{
  \nu^{\alpha}
  {\bf{v}}^{\alpha} \cdot {\bf{q}} }
 {  {\bf{v}}^{\alpha} \cdot {\bf{q}} - i \omega_{m} }
 + \nu^{\alpha} A^{\alpha}_{q} \right]
 \nonumber
 \\
 & = & \Pi_{0} ( q ) + \frac{1}{\beta V }
 \sum_{k}
 \left\{     G_{0} ( k) \Sigma_{F}^{(1)} ( k )  G_{0} ( k ) [ G_{0} ( k + q )
+ G_{0} ( k- q ) ]
 \right.
 \nonumber
 \\
 & & +
 \left. \frac{1}{2}
  G_{0} ( k )   [ \Gamma_{F}^{(1) } ( k , q )  G_{0} ( k + q )
  + \Gamma_{F}^{(1)} ( k , -q ) G_{0}  ( k -q ) ]
 \right\}
 \; \; \; ,
 \label{eq:Pipropres1}
 \end{eqnarray}
see Eqs.\ref{eq:Aalphaqres}-\ref{eq:gammaF1}.
In the three dimensional Coulomb-problem, the correction term in
Eq.\ref{eq:Pipropres1} has been
discussed in Refs.\cite{Geldart70,Holas79}. Note, however, that these authors
evaluate the
fermionic self-energy $\Sigma_{F}^{(1)} ( k )$ and vertex correction
$\Gamma^{(1)}_{F} ( k ,q )$ with the bare Coulomb interaction.
Holas {\it{et al.}}\cite{Holas79} find that the expansion in
Eq.\ref{eq:Pipropres1}
leads to unphysical singularities in the dielectric function close to the
plasmon-poles.

In contrast to the direct expansion of  $\Pi^{\ast} ( q )$ in powers
of the bare interaction,
in our method we first calculate the {\it{irreducible self-energy}}
of the effective bosonized Hamiltonian in powers of the
RPA-screened interaction, and then re-sum the
perturbation series via the Dyson equation.
The crucial point is that the problem of calculating corrections to the
RPA can be completely mapped onto an effective bosonic problem:
our functional bosonization method allows us
to explicitly construct
the {\it{interacting}} bosonic Hamiltonian.
Once we accept the validity of this mapping,
we can use standard many-body theory for bosonic systems,
which leaves us no choice: The corrections to the propagator
of this effective bosonic theory should be calculated by expanding its
irreducible self-energy
$\underline{\Sigma}^{\ast}_{q}$ (i.e. the {\it{inverse}} Greens-function)
in the number of internal bosonic loops,
and then solving the Dyson equation.
A similar re-summation has been suggested in
Refs.\cite{Holas79,Rajagopal72,Dharma76}, but
it is not so easy to justify this procedure at the fermionic level.
Our bosonization approach provides the natural justification for this
re-summation.
The unphysical singularities\cite{Holas79} that are encountered
in the direct perturbative approach might be understandable from the
point of view of bosonization:
they are artificially generated because one attempts to
calculate the bosonic single-particle Greens-function by direct expansion.
This expansion is bound to fail close to the poles of the Greens-function, i.e.
close to the plasmon poles.

\vspace{7mm}

Based on the insights gained from our bosonization approach,
we would like to suggest that
corrections to the RPA should be calculated by
expanding the generalized local field corrections
$\underline{g} ( q )$ in powers of the RPA-interaction
and then solving the bosonic Dyson equation \ref{eq:Piast2}.
We suspect that in this way
unphysical singularities in the dielectric function can be avoided.
In the high-density limit we obtain from the first line in Eq.\ref{eq:Dysonexp}
for the total proper polarization
 \begin{equation}
 \Pi^{\ast} ( q ) = \sum_{\alpha}
 \frac{
 \frac{ \nu^{\alpha} }{ 1 - A^{\alpha}_{q} }
  {\bf{v}}^{\alpha} \cdot {\bf{q}} }
 {  {\bf{v}}^{\alpha} \cdot {\bf{q}} - i \omega_{m} ( 1 - {B}^{\alpha}_{q} )
 - \frac{ ( i \omega_{m} )^2 }{  {\bf{v}}^{\alpha} \cdot {\bf{q}} }
{{B}}^{\alpha}_{q} }
 \; \; \; ,
 \label{eq:Pipropres}
 \end{equation}
with
 \begin{equation}
 {B}^{\alpha}_{q} = \frac{ A^{\alpha}_{q} }{ 1 - A^{\alpha}_{q} }
 \; \; \; .
 \label{eq:tildeAalphaq}
 \end{equation}
A detailed analysis of Eq.\ref{eq:Pipropres} and the resulting
dielectric function $\epsilon ( q ) = 1 + f_{\bf{q}} \Pi^{\ast} ( q )$ in
various dimensions
requires a careful analysis of the analytic properties of the function
$A^{\alpha}_{q}$ and extensive numerical
work, which will be presented elsewhere.
Here, we would like to restrict ourselves
to the discussion of the compressibility $K_{\rho}$,
which can be obtained as the "${\bf{q}} $-limit" of the exact
proper polarization (compressibility sum rule)\cite{Mahan81},
 \begin{equation}
 K_{\rho} = \lim_{ {\bf{q}} \rightarrow 0}  \left[
 \lim_{\omega_{m} \rightarrow 0 } \Pi^{\ast} ( {\bf{q}} , i \omega_{m} )
\right]
 \label{eq:compdef}
 \; \; \; .
 \end{equation}
As already mentioned, for sufficiently smooth Fermi surfaces the
dependence of $A^{\alpha}_{q}$ on the patch-index $\alpha$ can be ignored.
Setting $A_{0} = \lim_{ {\bf{q}} \rightarrow 0} [ \lim_{\omega_{m} \rightarrow
0 } A^{\alpha}_{q} ]$, we
obtain from Eq.\ref{eq:Pipropres}
 \begin{equation}
 K_{\rho} = \frac{ \nu}{1 -  A_{0}}
 \; \; \; .
 \label{eq:Krhores}
 \end{equation}
If we combine this expression with the estimate for $A_{0}$ given
in Eq.\ref{eq:Acb2}, we obtain
 \begin{equation}
 K_{\rho} = \frac{ \nu}{1 -  \gamma \left( \frac{\kappa}{k_{F}} \right)^2}
 \; \; \; ,
 \label{eq:Krhores2}
 \end{equation}
where $\gamma$ is a numerical constant of the order of unity.
Eq.\ref{eq:Krhores} agrees with the expression given by
Singwi {\it{et al.}}\cite{Singwi68}, who present also numerical
results for $\gamma$ as function of the density
of the electron gas.
Note that according to Eq.\ref{eq:Krhores} the
compressibility diverges and the system becomes unstable for $A_{0} \rightarrow
1$.
Eq.\ref{eq:Krhores} should be compared
with the expression for the compressibility
resulting from
approximation \ref{eq:Pipropres1},
$K_{\rho} \approx \nu ( 1 + A_{0} )$.
This is
the leading term in the expansion of Eq.\ref{eq:Krhores}
and does not predict any instabilities.

\section{Conclusions}
\label{sec:conclusions}

\setcounter{equation}{0}

The main results of this work can be summarized as follows:

(1) We have developed a functional integral method which allows us
to bosonize interating fermions in arbitrary dimensions.
In general, bosonization maps a system of interacting fermions onto an
effective system of
interacting bosons. The corrections to the non-interacting boson approximation
can be
calculated in a systematic way.

(2) There exists a physically interesting limit where the Gaussian
approximation for the
effective bosonic Hamiltonian becomes exact. This limit is characterized by the
requirements that the {\it{diagonal-patch approximation}} $(A1)$ and the
{\it{local linearization}} $(A2)$ are correct.
If the approximations $(A1)$ and $(A2)$ are made and the
radial cutoff $\lambda$ is removed,
the resulting interacting
Fermi system is exactly solvable in any dimension due to
an exact cancellation of self-energy and vertex corrections (generalized closed
loop theorem).
In the one-dimensional Tomonaga-Luttinger model the condition
$(A2)$ is satisfied by definition, while
$(A1)$ is satisfied as long as the two Fermi points are sufficiently separated,
so that
they cannot be connected
by the maximal momentum transfer of the interaction.
Because for realistic Fermi surfaces in $d>1$ there exist always
neighboring patches that can be connected by arbitrarily small momentum
transfers,
$(A1)$ is only approximately correct in higher dimensions.

(3) If the diagonal-patch approximation $(A1)$ is made but the energy
dispersion is
{\it{not}} linearized, we have explicitly calculated the leading correction to
the
Gaussian approximation, and have shown that the effect of interactions between
the bosons is negligible as long as the parameter
 $\left( \frac{ q_{c}}{k_{F} } \right)^{d} | {F}^{RPA}_{0} |  | C^{\alpha}| $
is small compared with unity. The dimensionless quantity $C^{\alpha}$ is
proportional to the derivative of the density of states with respect to the
chemical potential, and vanishes if the energy dispersion is linearized.
The origin  of an additional small parameter $C^{\alpha}$
in the correction to the RPA lies in the cancellation between
self-energy and vertex corrections, as described by the generalized
closed loop theorem.

(4) Our bosonization approach
maps the calculation of the
density-density correlation function of the Fermi system
on the problem of calculating the single-particle Greens-function
of an effective bosonic model. This mapping
explains the origin of unphysical singularities that arise in direct
perturbative expansions of the proper polarization, and
strongly suggests that corrections to the RPA should be calculated
by expanding the generalized local field corrections
in powers of the RPA-interaction.

\vspace{7mm}

The present work is the first step to go beyond the non-interacting boson
approximation.
We have calculated the bosonized Hamiltonian
as well as the density-density correlation function beyond the
Gaussian approximation.
The problem of calculating the non-Gaussian corrections to the
leading bosonization expression for the
single-particle Greens-function remains open.
This calculation is more difficult, because
it involves the  solution of a partial differential
equation\cite{Kopietz95,Kopietz94,Lee88,Schwinger62}.
If the energy dispersion is linearized, $\xi^{\alpha}_{\bf{q}} \approx
{\bf{v}}^{\alpha} \cdot {\bf{q}}
+ O ( {\bf{q}}^2 )$,
this differential equation is linear and first
order, and can be solved exactly.
Inclusion of non-linear terms of order ${\bf{q}}^2$ and higher in the
expansion of $\xi^{\alpha}_{\bf{q}}$
lead to higher spatial derivatives, which are
not so easy to handle.
Following the method used by Khveshchenko and Stamp\cite{Khveshchenko93},
it might be possible to set up some kind of
perturbation theory for the solution of this higher order partial
differential equation, but so far we
have not pursued this possibility.

It is tempting to take the
form of the Greens-function within the Gaussian approximation as a basis to
speculate on the effect of the non-Gaussian terms.
Within the Gaussian approximation
the general bosonization result for the real-space imaginary-time
single-particle Greens-function  associated with patch $\alpha$ can be written
as\cite{Kopietz94}
 \begin{equation}
 G^{\alpha} ( {\bf{r}} , \tau  )
   =
 G^{\alpha}_{0} ( {\bf{r}} , \tau  )
 e^{
 Q^{\alpha}
 ( {\bf{r}}  , \tau  ) }
 \label{eq:Gbos}
 \; \; \; ,
 \end{equation}
where $G^{\alpha}_{0} ( {\bf{r}} , \tau )$ is the non-interacting
Greens-function, and
 \begin{eqnarray}
 Q^{\alpha}
 ( {\bf{r}} , \tau ) & = &
 R^{\alpha}
 - S^{\alpha} ( {\bf{r}} , \tau )
  \; \;   , \;  \;
 R^{\alpha} =  \lim_{ {\bf{r}} , \tau \rightarrow 0} S^{\alpha} ({\bf{r}} ,
\tau )
 \;  ,
 \label{eq:Qdef}
 \\
 S^{\alpha}
 ( {\bf{r}}  , \tau   )
  & = &
 \frac{1}{\beta {{V}}} \sum_{ q }
 \frac{ \left[ \underline{f}^{RPA}_{q} \right]^{\alpha \alpha}
  \cos ( {\bf{q}} \cdot  {\bf{r}}
  - {\omega}_{m}  \tau  )
 }
 {
 ( i \omega_{m} -  {\bf{v}}^{\alpha} \cdot {\bf{q}}  )^{2 }}
  \label{eq:Sdef}
 \; \; \; .
 \end{eqnarray}
Note that this expression
depends exclusively on the RPA-screened interaction matrix
$\underline{f}^{RPA}_{q}$, as defined in Eq.\ref{eq:frpadef}.
Due to the linearization of the energy dispersion the
non-interacting real-space Greens-function $G^{\alpha}_{0} ( {\bf{r}} , \tau )$
is proportional to
$\delta^{(d-1)} ( {\bf{r}}_{\bot} )$, where ${\bf{r}}_{\bot}$ is the
$d-1$-dimensional projection
of ${\bf{r}}$ perpendicular to ${\bf{v}}^{\alpha}$. Therefore
at the level of the Gaussian approximation
we may replace
 \begin{equation}
 Q^{\alpha}
 ( {\bf{r}}  , \tau   )
 \rightarrow
 Q^{\alpha}
 (
 r_{\|} \hat{\bf{v}}^{\alpha} , \tau )
 \label{eq:Sreplace}
 \end{equation}
in Eq.\ref{eq:Gbos}, where $\hat{\bf{v}}^{\alpha}$ is a unit vector in the
direction of ${\bf{v}}^{\alpha}$ and $r_{\|} = {\bf{r}} \cdot
\hat{\bf{v}}^{\alpha}$.
If the non-linear terms in the energy dispersion
are taken into account,
we expect that the RPA-interaction is replaced by the exact effective
interaction, which takes
the local field corrections to the bare polarization bubbles into account.
Furthermore, the term $ {\bf{v}}^{\alpha} \cdot {\bf{q}} $ in the denominator
of Eq.\ref{eq:Sdef}
should be replaced by the
full dispersion $\xi^{\alpha}_{\bf{q}}$, and
the substitution \ref{eq:Sreplace}
cannot be made any more.
However, non-Gaussian terms will certainly also give rise to further
modifications,
so that it is not clear if the Greens-function
retains at least approximately
the above simple structure.

Finally, it would be interesting to calculate the corrections to the
RPA by means of the method outlined in Sec.\ref{subsec:diagram}.
This method is different from a straight-forward expansion of the proper
polarization, because
it is based on the loop-wise calculation of generalized local field corrections
associated with the patches.
The calculation of corrections to the
RPA has been discussed intensely in the $70$'s, but today the
interest in this problem has waned.
Our bosonization approach
sheds new light on the various approximations that can be found in the
literature,
and perhaps will revive the interest in this problem.

\section*{Acknowledgements}
\vspace{0.2cm}
We are grateful to Lorentz Bartosch for checking
a large part of the algebra presented in this work and
for suggesting some useful modifications.
We would also like to thank Walter Metzner
for discussions, for giving us
insight into his unpublished notes, and
for very useful comments on the manuscript.
Finally, we would like to thank
Volker Meden for his critical reading of the manuscript.

%
%
%

\begin{figure}
\caption{Graph of a squat box
$K^{\alpha}_{\Lambda , \lambda}$ with patch cutoff $\Lambda$
and radial cutoff $\lambda$ in three dimensions.
The vector ${\bf{k}}^{\alpha}$ points on the Fermi surface to the center
of patch $\tilde{K}^{\alpha}_{\Lambda}$.
}
\label{fig:patch}
\end{figure}

\begin{figure}
\caption{The boxes
$K^{\alpha}_{  \Lambda }$ for a spherical Fermi surface in $d=2$.
The dashed lines mark the corresponding squat boxes
$K^{\alpha}_{\Lambda , \lambda}$
that contain only the degrees of freedom in the vicinity
of the Fermi surface.
}
\label{fig:patch2}
\end{figure}

\begin{figure}
\caption{Feynman-diagram representing $Tr[ \hat{G}_{0} \hat{V} ]^{n}$, see
Eqs.\ref{eq:tracelogexp} and
\ref{eq:Seffphin}.
The lines with arrows denote fermionic Greens-functions, while the
wavy lines denote external $\phi^{\alpha}$-fields.}
\label{fig:closedloop}
\end{figure}

\begin{figure}
\caption{
Leading  contributions to the bosonic self-energy, see Eq.\ref{eq:self1}.
The dark circle denotes the vertex $\Gamma_{2}^{(1)}$ defined in
Eq.\ref{eq:Gamma21def} and the shaded square
represents the vertex $\Gamma_{4}^{(0)}$ defined in Eq.\ref{eq:Gamma04}.
The dashed arrows denote the collective density field
$\tilde{\rho}^{\alpha}$, and the dashed loop is the
Gaussian propagator of the $\tilde{\rho}^{\alpha}$-field, see
Eq.\ref{eq:rhopropgauss}.}
\label{fig:self1}
\end{figure}

\begin{figure}
\caption{
Leading local-field corrections to the polarization bubble.
(a) und (b) are the leading self-energy corrections (see Eq.\ref{eq:sigmaF1}),
while
(c) is the leading vertex correction (see Eq.\ref{eq:gammaF1}).
The thick wavy line denotes the RPA-interaction, as defined in (d).
The thin wavy line in (d) represents the bare interaction.}
\label{fig:bubblecor}
\end{figure}

\end{document}